\documentclass[12pt]{article}

\usepackage{amsmath,amsfonts,amssymb}
\usepackage[nosort]{cite}
\usepackage{hyperref}
\usepackage{graphicx}
\usepackage{color}
\usepackage{epsfig}
\usepackage{psfrag}	
\usepackage{tikz}
\usepackage{slashed}
\usepackage{ulem}
\usepackage[font=small,labelfont=bf,width=1.1\textwidth]{caption}
\usepackage{comment}
\usepackage[all]{xy}
\usepackage{subcaption}			

\numberwithin{equation}{section}
\usepackage{color}

\newcommand \be{\begin{eqnarray}}
\newcommand \ee{\end{eqnarray}}

\usetikzlibrary{decorations.pathreplacing,calc}

\DeclareMathOperator{\tr}{tr}

\DeclareMathOperator{\vol}{vol}
\DeclareMathOperator{\diag}{diag}

\def\bO {\mathbb{O}}

\def\bQ {\mathbb{Q}}

\def\aP{{\mathsf{P}}}
\def\aM{{\mathsf{M}}}
\def\aK{{\mathsf{K}}}
\def\aD{{\mathsf{D}}}
\def\aR{{\mathsf{R}}}
\def\aQ{{\mathsf{Q}}}
\def\aS{{\mathsf{S}}}
\def\aU{{\mathsf{U}}}
\def\aT{{\mathsf{T}}}

\newcommand{\bea}{\begin{eqnarray}}
\newcommand{\eea}{\end{eqnarray}}
\newcommand{\beq}{\begin{equation}}
\newcommand{\eeq}{\end{equation}}
\newcommand{\bal}{\begin{equation}\begin{aligned}}
\newcommand{\eal}{\end{aligned} \end{equation}}

\newcommand{\vev}[1]{{\left< {#1} \right>}}
\newcommand{\bra}[1]{{\left< {#1} \right|}}
\newcommand{\ket}[1]{{\left| {#1} \right>}}

\newcommand{\address}[1]{\vbox{\center\em#1}}
\renewcommand{\title}[1]{\vbox{\center\huge{#1}}\vspace{5mm}}

\newcommand{\bD}{{\mathbb D}}
\newcommand{\cA}{{\mathcal A}}

\newcommand{\cN}{{\mathcal N}}
\newcommand{\cP}{{\mathcal P}}

\newcommand{\cO}{{\mathcal O}}

\newcommand{\su}{\mathfrak{su}}
\newcommand{\so}{\mathfrak{so}}
\renewcommand{\sl}{\mathfrak{sl}}

\newcommand{\osp}{\mathfrak{osp}}
\newcommand{\sof}{\mathfrak{so}}

\newcommand{\im}{i}

\newcommand{\diff}{d}
\newcommand{\fin}{\text{finite}}
\newcommand{\RicciScalar}{R}

\usepackage{romanbar} 
\newcommand{\IIFundForm}{\Romanbar{2}}
\usepackage{tensor} 
\newcommand{\norm}[1]{\left| #1 \right|}
\newcommand{\ho}[2][]{\mathcal{O}(#2^{#1})}

\usepackage{stmaryrd} 

\begin{document}

\begin{titlepage}
\begin{center}

\vspace*{20mm}

\renewcommand{\thefootnote}{$\alph{footnote}$}
\vspace{10mm}

\title{Defect CFT techniques in the 6d $\mathcal{N} = (2,0)$ theory}
Nadav Drukker,\footnote{\href{mailto:nadav.drukker@gmail.com}{nadav.drukker@gmail.com}}
Malte Probst,\footnote{\href{mailto:mltprbst@gmail.com}{mltprbst@gmail.com}}
and 
Maxime Tr\'epanier\footnote{\href{mailto:trepanier.maxime@gmail.com}{trepanier.maxime@gmail.com}}
\address{Department of Mathematics, King's College London,\\The Strand, London WC2R
2LS, United Kingdom}

\renewcommand{\thefootnote}{\arabic{footnote}}
\setcounter{footnote}{0}

\end{center}

\vspace{8mm}
\abstract{
\normalsize{
\noindent
Surface operators are among the most important observables of the 6d
$\mathcal{N} = (2,0)$ theory. Here we apply the tools of defect CFT to study
local operator insertions into the $1/2$-BPS plane. We first relate the 2-point
function of the displacement operator to the expectation value of the bulk
stress tensor and translate this relation into a constraint on the
anomaly coefficients associated with the defect. Secondly, we study the defect 
operator expansion of the stress tensor multiplet and 
identify several new operators of the defect CFT. Technical results derived along 
the way include the explicit supersymmetry tranformations of the stress tensor multiplet 
and the classification of unitary representations of the superconformal algebra preserved 
by the defect. 
}}
\vfill
\end{titlepage}

\tableofcontents

\section{Introduction}

A natural class of observables of the $\cN = (2,0)$ 6d SCFT is that of surface
operators~\cite{ganor:1996nf}. These operators share many properties with the
much studied Wilson loops of gauge theories: they are extended objects which can
enjoy superconformal symmetry~\cite{Howe:1997ue}, in some cases have a holographic
description~\cite{maldacena:1998im}, and in the abelian $\cN = (2,0)$ theory
admit a field realisation as the integral of the 2-form $B$ field, akin to
a gauge connection~\cite{henningson:1999xi, hen99, 
gus03,gustavsson:2004gj,Drukker:2020dcz}. These
similarities suggest that some of the methods which have proven successful in the 
study of Wilson loops can be applied to the $\cN = (2,0)$ theory as well, providing 
a window into its dynamics.

In this paper we apply the framework of defect CFT to the surface operators of
the $\cN = (2,0)$ theory. We adopt the approach of the conformal bootstrap
program~\cite{Ferrara:1973yt,Ferrara:1973vz,Polyakov:1974gs,Rattazzi:2008pe,Liendo:2012hy,Beem:2015aoa}
and use the symmetries preserved by the surface operators to constrain their
correlators with other bulk operators as well as local operator insertions on the
surface. One of the virtues of this description is that it does not rely on a field 
realisation and therefore is applicable to the nonabelian theory. 

We focus on 1/2-BPS defects because they preserve the largest amount of symmetry. These
are surface operators defined over a plane and expected to be labeled by a
representation of the $ADE$ group of the $\cN = (2,0)$
theory~\cite{witten:1995zh,DHoker:2008rje, bachas:2013vza}. We consider local
operator insertions into the defect, the simplest example encoding an infinitesimal 
geometric deformation of the plane itself. Because the plane preserves 
superconformal symmetry, the correlators of local operator insertions are constrained 
and obey the axioms of a dCFT---the 2- and 3-point functions are fixed up to a small set
of numbers defining the dCFT, which make up the dCFT data.

Explicitly, consider a correlator involving such a surface operator $V$. While translating 
the plane along parallel directions leaves the correlator invariant, translations in directions 
transverse to the plane do not. Instead, the stress tensor receives a contribution from a 
contact term localised on the defect (at $x = 0$):
\begin{align}
  \partial_\mu T^{\mu m}(\sigma,x) \, V &= V[\bD^{m}(\sigma)] \delta^{(4)}(x).
  \label{eqn:dispop}
\end{align}
The index $\mu = 1,\dots,6$ runs over all spacetime coordinates, while $m=1,\dots,4$
are the coordinates transverse to the plane.
We use the notation $V[\hat{\mathcal{O}}(\sigma)]$ to denote the planar surface
operator with a defect operator $\hat{\mathcal{O}}$ inserted at a point $\sigma$ on the
plane.

Equation~\eqref{eqn:dispop} is an operator equation, so it holds inside
correlation functions.  It defines $\bD$, known as the displacement
operator.  In addition, because $V$ preserves some supersymmetries, the
displacement operator sits in a multiplet containing also contact terms for the
divergence of the broken super- and R-current, which we label $\bQ$ and $\bO$, 
respectively.

It turns out that these defect operators enjoy a very favorable position:
not only are they highly constrained by the residual symmetry (which includes
the 2d rigid superconformal symmetry), but they also correspond to
interesting physical quantities~\cite{Lewkowycz:2014jia,bianchi:2015liz}.%
\footnote{
  Note that the dCFT is not expected to contain a conserved stress
  tensor~\cite{Herzog:2017xha} and the rigid conformal symmetry is not
  necessarily enhanced to Virasoro symmetry.
}
Indeed it is easy to show that, as a consequence of~\eqref{eqn:dispop}, 
the insertion of a displacement operator $\bD$ corresponds to small deformations 
of the plane, and thus captures the shape dependence of surface operators.

This paper revolves around two correlators that capture 
physical properties of the defect. The first one is the 2-point function of
displacement operators. Using the residual conformal symmetry of the plane and
reading the conformal dimension $\Delta_\bD = 3$ from~\eqref{eqn:dispop}, the
2-point function is constrained up to a single coefficient $C_\bD$ to be
\begin{align}
  \vev{V[\bD^{m}(\sigma) \bD^{n}(0)]} =
  \frac{C_\bD \delta^{mn}}{\pi^2 |\sigma|^6}\,.
  \label{eqn:disp2pts}
\end{align}
Notice however that, unlike most operators, the normalisation of $\bD$ is
already fixed by the normalisation of $T^{\mu\nu}$ from~\eqref{eqn:dispop}, so
that $C_\bD$ is part of the data characterising the dCFT~\cite{correa:2012at} 
(the factor $\pi^2$ is for convenience).

The second is the stress tensor, which in the presence of the defect acquires an
expectation value. Both the components of the tensor along the defect
$T^{ab}$ and orthogonal to it $T^{mn}$ can have a nonzero 1-point function, and
they are fixed by conformal invariance up to an arbitrary coefficient
$h_T$ to be
\begin{align}
  \vev{T^{ab}(\sigma, x) V} = \frac{h_T \eta^{ab}}{\pi^3 x^6}\,, \qquad
  \vev{T^{mn}(\sigma, x) V} = -\frac{h_T (\delta^{mn}-2 x^{m} x^{n}/x^2)}
  {\pi^3 x^6}\,.
  \label{eqn:stresstensor1pt}
\end{align}
$T(\sigma,x)$ is inserted at a distance $x$ from the defect, and obviously the
correlators do not depend on the coordinate $\sigma$ by translation invariance
along the plane.  $\eta^{ab} = \diag{(-1,1)}$ is the Minkowski metric.

In theories with only conformal invariance the coefficients $h_T$
and $C_\bD$ are independent quantities~\cite{Billo:2016cpy}, but in theories
with enough supersymmetries one can use superconformal Ward identities to relate
them~\cite{bianchi:2018zpb}. For our surface operators we show in
Section~\ref{sec:BulkDefectCorrelators} that
\begin{align}
  h_T = \frac{3 C_\bD}{80}\,.
  \label{eqn:result}
\end{align}
To derive this result, we obtain the transformations of the stress tensor
multiplet under supersymmetry~\eqref{eqn:susyTmult}, which is also an important
result of Section~\ref{sec:BulkDefectCorrelators}.

Analogous relations between $h_T$ and $C_\bD$ were first derived using the same
techniques for the 1/2-BPS Wilson loops of 4d $\cN = 2$
theories~\cite{bianchi:2018zpb} and the 1/6-BPS bosonic loops of
ABJM~\cite{Bianchi:2018scb}, proving the conjecture
of~\cite{Lewkowycz:2013laa,Fiol:2015spa}. A similar analysis was also applied
recently to surface operators in 4d $\cN = 1$ theories~\cite{Bianchi:2019sxz}.  
All these different examples show how the language of dCFT is a powerful and
universal tool to study superconformal defects.

More than simply equating different constants, the relation~\eqref{eqn:result}
has an important physical consequence.  Recall that surface operators in CFTs
typically have a conformal anomaly, which manifests itself as a divergence in
the expectation value. The anomaly density $\cA_\Sigma$ is the sum of
conformal invariants~\cite{deser:1993yx,Boulanger:2007st} and can be written as
\begin{align}
  \mathcal{A}_\Sigma = \frac{1}{4 \pi} \left[
  a \RicciScalar^{\Sigma} + b_1 \tr{\tilde{\IIFundForm}^2} + b_2 \tr{W} +c(\partial n)^2
  \right],
  \label{eqn:anomaly}
\end{align}
where the invariants
$\RicciScalar^{\Sigma}, \tr{\tilde{\IIFundForm}^2}, \tr{W},(\partial n)^2$
are local quantities depending on the embedding of the surface (see
Appendix~\ref{app:anomaly} for a review), while the coefficients $a, b_1, b_2$
and $c$ are known as anomaly coefficients and depend on the specifics
of the theory and operator in question.

In Section~\ref{sec:anomaly} we relate the coefficients $b_1$, $b_2$, $c$
to $C_\bD$, $h_T$ and an additional constant $C_\bO$ to be introduced
in~\eqref{eqn:DISP2pts}. In the language of anomaly coefficients, the
result~\eqref{eqn:result} along with the relative
normalisations~\eqref{eqn:cdco} of the operators in the displacement multiplet
can be stated as
\begin{align}
  c = -b_1/2\,, \qquad
  b_1 = -b_2\,.
  \label{eqn:anomalyresult}
\end{align}
We emphasize that these identities are a consequence of supersymmetry and hold
for any 1/2-BPS operator of the $\cN = (2,0)$ theory and for any $ADE$ group. In
particular, the second identity agrees with the explicit holographic
calculations of~\cite{graham:1999pm,Drukker:2020dcz,Drukker:2020swu} and was
conjectured to come from supersymmetry in~\cite{Bianchi:2019sxz}. The two remaining
anomaly coefficients $a$ and $b_1$ were calculated at $N=1$
in~\cite{Drukker:2020dcz} and for $N>1$ using holographic entanglement entropy
in the presence of surface
operators~\cite{Gentle:2015jma,Rodgers:2018mvq,Jensen:2018rxu,Estes:2018tnu},
and the superconformal index~\cite{Chalabi:2020iie}. 

Finally, in Section~\ref{sec:doe} we expand our scope and consider the analog of
the operator product expansion but for bulk operators in the presence of a
defect---the defect operator expansion (dOE)~\cite{Diehl:1981jgg,Cardy:1991tv}.
This expansion gives a representation of bulk operators near the defect in terms
of insertions of defect operators. To understand what these defect operators are 
more generally, we classify unitary multiplets of the algebra preserved by the
defect. We then look at operators in the stress tensor multiplet and determine
the short multiplets arising in their dOE. We find a new marginal defect
operator, which we associate with the RG flow between the nonsupersymmetric and
1/2-BPS surface operator discussed in~\cite{Drukker:2020swu}.\footnote{%
  This is analogous to the flow of Wilson line operators introduced
  in~\cite{Polchinski:2011im,Alday:2007he}.
}

In addition to this result, we find that the defect operator
expansion provides a useful framework and makes the constraints imposed by the
preserved symmetries manifest. In fact, in Section~\ref{sec:dOEsusy} we use the
dOE and representation theory to give a different perspective on the
relation~\eqref{eqn:result}. Unlike in Section~\ref{sec:BulkDefectCorrelators},
where~\eqref{eqn:result} follows from a technical calculation, we are able to
conclude directly that $h_T$ and $C_\bD$ must be related. This suggests a
strategy for determining the minimal amount of supersymmetry required in order 
for the conjecture of~\cite{Lewkowycz:2013laa}, which relates these coefficients 
in the case of supersymmetric Wilson loops, to hold (see also~\cite{bianchi:2015liz} and
references therein for a similar conjecture in the context of entanglement entropy).

Some auxiliary results are collected in appendices.
Appendix~\ref{sec:convention} summarises our conventions and the gamma matrices
used throughout the paper. Appendix~\ref{app:anomaly} provides a short review of
the Weyl anomaly for surface operators. In
Appendix~\ref{app:ConformalWardIdentities} we show how to constrain correlators
containing both bulk and defect operators using conformal symmetry.
Appendix~\ref{sec:ospalgebra} reviews the 2 algebras used in this paper: the
$\osp(8^*|4)$ symmetry of the bulk theory and the $\osp(4^*|2) \oplus
\osp(4^*|2)$ symmetry preserved by the defect.

{\textit{Note added:} In the last stages of preparation of this paper, the
classification of unitary multiplets of $\osp(4^*|2)$ presented in
Section~\ref{sec:unitarity} appeared in~\cite{Agmon:2020pde}.}

\section{Displacement multiplet}
\label{sec:disp}

As far as defect operators go, the displacement operator is pretty universal.
As~\eqref{eqn:dispop} suggests, any defect breaking translation symmetry
contains that defect operator. For this reason, it has appeared in many
contexts: the prototypical example is the 1/2-BPS Wilson line in $\cN = 4$ SYM,
where the study of deformations and operator insertions was initiated
in~\cite{Drukker:2006xg}, but many other examples have been studied over the
years and follow the general analysis of~\cite{Billo:2016cpy}.

In the case of $\cN = (2,0)$, we are mostly interested in the multiplet which
contains the displacement operator.  Of the full superconformal algebra
$\osp(8^\ast|4)$, the 1/2-BPS plane preserves a 2d conformal algebra
$\sof(2,2)_\parallel$ in the directions parallel to the plane, along with
rotations of the transverse directions $\sof(4)_\perp$ and an $\sof(4)_R$
R-symmetry. In addition, it also preserves half the supersymmetries $\aQ_+$ (and
$\bar{\aS}_+$) such that $\aQ_+ V = 0$. These are obtained by a half-rank
projector $\aQ_+ = \Pi_+ \aQ$ whose explicit definition can be found
in~\eqref{eqn:ospprojector}.  The preserved generators form an $\osp(4^*|2)
\oplus \osp(4^*|2)$ subalgebra~\cite{dhoker:2008wvd}, detailed in
Appendix~\ref{sec:ospsubalg}.

Importantly, in direct analogy to~\eqref{eqn:dispop}, the Ward identities 
associated to the remaining broken super- and R-symmetries also receive contributions 
localised on the defect, which give rise to defect operators $\bQ$ and $\bO^{i}$, 
encoding the nontrivial response of the defect to the broken generators.
Explicitly, the conservation laws associated with the R-current $j$ and 
the supercurrent $J$ are broken as follows:
\begin{equation}
  \begin{aligned}
    \partial_\mu T^{\mu m} V &= V[\bD^{m}] \delta^{(4)}(x)\,, \\
    \partial_\mu (\Pi_- J^{\mu}) V &= V[\bQ] \delta^{(4)}(x)\,, \\
    \partial_\mu j^{\mu i 5} V &= V[\bO^{i}] \delta^{(4)}(x)\,.
  \end{aligned}
  \label{eqn:broken}
\end{equation}
In this equation, $i=1,\dots,4$ is the R-symmetry index of $\sof(4)_R$. The
spinor indices of $J^\mu_{\alpha \check{\alpha}}$ and 
$\bQ_{\alpha \check{\alpha}}$ are suppressed and follow the conventions outlined
in appendix~\ref{sec:convention} (see however footnote~\ref{fn:repJ}). For the 
definition of $\Pi_-$, see~\eqref{eqn:ospprojector}.

As mentioned previously, the (nonabelian) theory does not have a known field
realisation, so we cannot write these operators in terms of fundamental fields.
We can however derive some of their properties purely from representation
theory. The full multiplet as derived in Appendix~\ref{sec:repderivation} reads
\begin{gather}
  \label{eqn:dispsusy}
  \begin{split}
    \delta_+ \bD_{m} &=
    \frac{1}{2} \varepsilon_+ \gamma_{a m} \partial^a \bQ\,,\\
    \delta_+ \bQ &=
      2 \varepsilon_+ \gamma_{m} \bD^m
    - 2 \varepsilon_+ \gamma_a \check{\gamma}_{i5}
    \partial^a \bO^{i}\,,\\
    \delta_+ \bO_{i} &=
    -\frac{1}{2} \varepsilon_+ \check{\gamma}_{i5} \bQ\,.
  \end{split}
\end{gather}
$\delta_+ = \varepsilon_+ \aQ_+$ is a variation with respect to
the preserved supercharges and $\varepsilon_+ = \varepsilon_+ \Pi_+$.

\subsection{Superconformal Ward identity}

The 2-point functions of these operators is easy to find. Both $\bD$ and $\bO$
transform as scalars with respect to the 2d conformal symmetry, while $\bQ$ is a
spinor. Their conformal dimensions can also be read from~\eqref{eqn:broken} and
are $\Delta_\bD = 3$, $\Delta_\bQ = 5/2$ and $\Delta_\bO = 2$. Consequently,
using the preserved bosonic symmetries, their 2-point functions are (up to some
arbitrary coefficients $C_\bD$, $C_\bQ$, $C_\bO$)
\beq
\begin{split}
  \vev{V[\bD^m(\sigma) \bD^n(0)]} &=
  \frac{C_\bD \delta^{mn}}{\pi^2
    \norm{\sigma}^{6}}\,, \\
    \vev{V[\bQ(\sigma) \bQ(0)]} &=
    \frac{C_\bQ \left( \gamma_a \sigma^a \Pi_- \right)}{\pi^2
    \norm{\sigma}^{6}}\,, \\
  \vev{V[\bO_{i}(\sigma) \bO_{j}(0)]} &=
  \frac{C_\bO \delta_{ij}}{ \pi^2 \norm{\sigma}^{4}}\,.
\end{split}
  \label{eqn:DISP2pts}
\eeq

As $\bQ$ is a 2d spinor, its 2-point function should be written in terms of the corresponding 
2d gamma matrices. In order to emphasize the relation between the respective symmetry algebras in 6d 
and 2d, we write these matrices as blocks of their 6d counterparts obtained by the projector $\Pi_-$.

We can now relate $C_\bO$ and $C_\bQ$ to $C_\bD$ using superconformal Ward
identities associated to the preserved supersymmetries. Apply the supersymmetry 
transformations~\eqref{eqn:dispsusy} to the vanishing correlator $\vev{V[\bQ_{\beta \check\beta} \bO_i]}$
to find
\beq
  -\frac{1}{2} \tensor{\left( \check{\gamma}_{i 5}
  \right)}{_{\check{\alpha}}^{\check{\gamma}}}
  \vev{V[\bQ_{\beta \check{\beta}}
  \bQ_{\alpha \check{\gamma}}]}
  = 2 \left( \gamma_a \check{\gamma}_{j 5} \Pi_- c \Omega \right)
  _{\alpha \beta \check{\alpha} \check{\beta}}
  \partial^a \vev{V[\bO^{j} \bO_{i}]}.
\eeq
Substituting the explicit 2-point functions~\eqref{eqn:DISP2pts}, we obtain the linear relation 
$C_\bQ = - 16 C_\bO$.  In the same fashion, the Ward identity associated to 
$\vev{V[\bQ_{\beta \check{\beta}} \bD_{m}]}$ leads to
\beq
  2 \left( \gamma_{n} \Pi_- c \Omega \right)
  _{\alpha \check{\alpha} \beta \check{\beta}}
  \vev{V[\bD^{n} \bD_{m}]}
  = -\frac{1}{2} \tensor{\left( \gamma_{a m} \right)}
  {_{\alpha}^{\gamma}}
  \partial^a \vev{V[\bQ_{\beta \check{\beta}}
  \bQ_{\gamma \check{\alpha}}]},
\eeq
which serves to relate $C_\bD$ to $C_\bQ$. Altogether, we find that the normalisations of the 
2-point functions obey
\begin{align}
  C_\bD = - C_\bQ = 16 C_\bO.
  \label{eqn:cdco}
\end{align}

\section{Stress tensor correlators}
\label{sec:BulkDefectCorrelators}

Some of the most important operators in any theory are the stress tensor and its
multiplet. In the presence of the 1/2-BPS defect, their expectation values are highly
constrained by the residual symmetry: typically the $\sof(2,2)_\parallel
\oplus \sof(4)_\perp \oplus \sof(4)_R$ bosonic subalgebra of preserved symmetries is 
powerful enough to fix them up a to a constant (see e.g.~\eqref{eqn:stresstensor1pt}).

In addition to the constraints imposed by conformal symmetry, supersymmetry relates
correlators of different operators in the same multiplet. Adapting the strategy
of~\cite{bianchi:2018zpb,Bianchi:2018scb,Bianchi:2019sxz},
the key to deriving~\eqref{eqn:result} is to focus on the correlator
$\vev{T^{\mu \nu}(x) V[\bD^m(\sigma)]}$, which is entirely fixed in terms of the
constants $C_\bD$ and $h_T$~\cite{Billo:2016cpy}. The kinematics of that
correlator admit 2 independent tensor structures with their own
coefficient. They are related to $C_\bD$ by taking
the divergence
\begin{align}
  \partial_\mu \vev{T^{\mu m} V[\bD^n]} = \vev{V[\bD^m \bD^n]} \propto C_\bD\,,
  \label{eqn:ExDivergence}
\end{align}
and to $h_T$ by integrating the displacement operator over the surface, which simply translates
the defect
\begin{align}
  \int_{\mathbb{R}^2} \diff^2 \sigma
  \vev{T^{\mu\nu}(0,x) V[\bD^m(\sigma)]} = \partial^m \vev{T^{\mu\nu}(0,x) V}
  \propto h_T\,.
  \label{eqn:ExIntegrated}
\end{align}
We stress that this does not provide in itself a relation between $C_\bD$ and
$h_T$, as can be checked using the explicit form of the correlators (see
equation (6.2) of~\cite{bianchi:2015liz}).

Instead, to derive the relation, we should use superconformal Ward identites to relate this correlator to
$\vev{O^{i5} V[\bO^{j}]}$, where $O$ is the superconformal primary of the stress
tensor multiplet. Because the latter admits only a single tensor structure, this would 
imply that $C_\bD$ and $h_T$ are related.

In order to derive this result, we need the explicit supersymmetry
transformations of the stress tensor multiplet, which are summarised
in~\eqref{eqn:susyTmult}. We also need the 1-point functions of the stress
tensor appearing on the right-hand side of~\eqref{eqn:ExIntegrated}, which are
derived in Section~\ref{sec:1PTstress} (the 2-point functions of the displacement
multiplet are given in~\eqref{eqn:DISP2pts}). Then, we use the
supersymmetric Ward identities associated with correlators of the form
$\langle \cO V[\hat\cO] \rangle$ to derive~\eqref{eqn:result}.

\subsection{Stress tensor multiplet}
\label{sec:stress}

We begin by obtaining explicit supersymmetry transformations for the stress
tensor multiplet, whose content is derived from representation
theory and can be found in~\cite{gunaydin:1984wc}, where it is presented as a
massless graviton multiplet (see also~\cite{cordova:2016emh,Buican:2016hpb} for
an overview of superconformal multiplets in various dimensions).

The primaries of any multiplet are labelled by their transformation under Lorentz
symmetry $[j_1,j_2,j_3]_{\mathfrak{su}(4)}$, R-symmetry
$(R_1, R_2)_{\mathfrak{sp}(2)}$ as well as their conformal dimension $\Delta$.\footnote{%
  These Dynkin labels are related to the usual $\so(1,5)$ and $\so(5)$ labels by
  \begin{align}
    [j_1,j_2,j_3]_{\mathfrak{su}(4)} = [j_2,j_1,j_3]_{\mathfrak{so}(1,5)}\,,
    \qquad
    (R_1, R_2)_{\mathfrak{sp}(2)} = (R_2, R_1)_{\mathfrak{so}(5)}\,.
\nonumber  \end{align}
}
In the notation of ~\cite{cordova:2016emh}, the stress tensor multiplet is the
$D_1[0,0,0]^{(0,2)}_4$ multiplet (with representations written
as $[j_1,j_2,j_3]^{(R_1,R_2)}_\Delta$). Its primaries are
\begin{itemize}
  \item $T^{\mu\nu}$, the stress tensor ($[0,2,0]^{(0,0)}_6 = \mathbf{20}$). It contains a null
    state, since $\partial_\mu T^{\mu\nu} = 0$, and has $20-6$ degrees of
    freedom.
  \item $J^\mu_{\alpha \check{\alpha}}$, the supercurrent
    ($[1,1,0]^{(1,0)}_{11/2} = \mathbf{20} \cdot \mathbf{4}$). It also has a null state
    $\partial_\mu J^\mu_{\alpha \check{\alpha}} = 0$, satisfies
    $\tensor{(\bar{\gamma}_\mu)}{_{\dot{\alpha}}^\beta} J^\mu_{\beta \check{\beta}} = 0$,
    and contains $80-16$ degrees of freedom.\footnote{\label{fn:repJ}Note that $J$ transforms in 
    the $[1,1,0]$ irrep. Since the tensor product of a vector and a chiral spinor decomposes into 
    $[1,1,0] \oplus [0,0,1]$, we can write $J$ with indices $\mu$ and $\alpha$, provided we project 
    out the antichiral spinor by requiring $\tensor{(\bar{\gamma}_\mu)}{_{\dot{\alpha}}^\beta} 
    J^\mu_{\beta \check{\beta}} = 0$.}
  \item $j^{\mu[IJ]}$, the R-current ($[0,1,0]^{(2,0)}_5 = \mathbf{6} \cdot
    \mathbf{10}$). It has a null state $\partial_\mu j^{\mu IJ} = 0$, and
    contains $60-10$ degrees of freedom.
  \item $H_{\mu\nu\rho}^I$, a self-dual 3-form ($[2,0,0]^{(0,1)}_5 = \mathbf{10}
    \cdot \mathbf{5}$) containing $50$ degrees of freedom.
  \item $\chi_{\alpha \check{\alpha}}^I$, a fermion ($[1,0,0]^{(1,1)}_{9/2} = \mathbf{4}
    \cdot \mathbf{16}$) satisfying $\tensor{(\check{\gamma}_I)}{_{\check{\alpha}}^{\check{\beta}}}
    \chi^I_{\beta \check{\beta}} = 0$ and containing $64$ degrees of freedom.
  \item $O^{(IJ)}$, a scalar ($[0,0,0]^{(0,2)}_4 = \mathbf{14}$) with 14 degrees
    of freedom. It is the superprimary of the multiplet.
\end{itemize}
Together with their descendants, these form an on-shell multiplet with $128$ bosonic operators 
(and a matching number of fermionic operators).

In addition to the operator content, we need below the explicit supersymmetry
transformations, which have not been calculated before to
the best of our knowledge. These can be obtained in a variety of ways (e.g.
oscillator constructions~\cite{gunaydin:1984wc} and superspace
transformations~\cite{Park:1998nra,Cordova:2019wns}), but here we simply
list the terms allowed by Lorentz and R-symmetry and fix the coefficients by requiring closure of the 
algebra, i.e. imposing that on every operator $\left\{ \aQ, \aQ \right\} \Phi = 2 \aP \Phi$. 
Importantly, imposing this condition is made easy because we already know the operator content.

We start from the superprimary $O^{IJ}$. Since $\aQ$ transforms as
$[1,0,0]^{(1,0)}_{1/2}$, we know from representation theory that the product
$\aQ O$ can contain
\begin{align}
  [1,0,0]^{(1,2)}_{9/2} \oplus [1,0,0]^{(1,1)}_{9/2}\,,
\end{align}
but as $[1,0,0]^{(1,2)}_{9/2}$ does not appear in the multiplet, we remove it. 
The remaining term $[1,0,0]^{(1,1)}_{9/2}$ can be constructed explicitly and is
fixed up to a constant $c_1$
\begin{align}
  \aQ_{\alpha \check{\alpha}} O^{IJ} =
  c_1 (\check{\gamma}^{(I} \chi^{J)})_{\alpha \check{\alpha}}\,.
\end{align}
The transformation of $\chi$ is more complicated but the same analysis leads to
\begin{equation}
  \begin{aligned}
    \aQ_{\alpha \check{\alpha}} \chi^I_{\beta \check{\beta}} &=
    c_2 (\gamma^{\mu\nu\rho})_{\alpha\beta}
    \left( \check{\gamma}^{IJ} + 4 \delta^{IJ} \right)_{\check{\alpha} \check{\beta}}
    H^J_{\mu\nu\rho} 
    + c_3 (\gamma_\mu)_{\alpha \beta} \left( \check{\gamma}^{IJK} + 
    3 \delta^{IJ} \check{\gamma}^{K} \right)_{\check{\alpha} \check{\beta}}
    j^{\mu}_{JK}\\
    & \quad + d_1 (\gamma^\mu)_{\alpha \beta}
    \left( \check{\gamma}^{J} \right)_{\check{\alpha} \check{\beta}} \partial_\mu
    O^{IJ}\,.
  \end{aligned}
  \label{eqn:qchi}
\end{equation}
It is easy to check that
\begin{align}
  \left\{ \aQ_{\alpha \check{\alpha}}, \aQ_{\beta \check{\beta}}
  \right\} O^{IJ} =
  2 c_1 d_1 (\gamma^\mu)_{\alpha \beta} \Omega_{\check{\alpha} \check{\beta}}
  \partial_\mu O^{IJ}\,,
  \label{eqn:jacobi}
\end{align}
so the algebra closes provided $c_1 d_1 = 1$ (we identify $\aP_\mu =
\partial_\mu$, see~\eqref{eqn:repconf}).

We can proceed this way for the full multiplet and build the supersymmetry 
transformations. Checking for closure of the algebra becomes a tedious (if
straightforward) task and is not very illuminating, so we omit the details. The
end result is (with $\delta =
\varepsilon^{\alpha \check{\alpha}} \aQ_{\alpha \check{\alpha}}$)
\\
\begin{minipage}[b]{0.21\textwidth}
  \begin{tikzpicture}[scale=2.4]
    \node (np) at (0,1) {$T^{\mu\nu}$};
    \node (nq) at (0,0.5) {$J^\mu_{\alpha \check{\alpha}}$};
    \node (nr) at (0.5,0) {$j^\mu_{IJ}$};
    \node (nm) at (-0.5,0) {$H^I_{\mu\nu\rho}$};
    \node (ns) at (0,-0.5) {$\chi^I_{\alpha \check{\alpha}}$};
    \node (nk) at (0,-1) {$O^{IJ}$};
    \draw[thick,->] (nk) edge node[right,anchor=west] {$\aQ$} (ns) (ns) edge (nr) (nr) edge (nq) (nq) edge (np)
    (ns) edge (nm) (nm) edge (nq);
  \end{tikzpicture}
  \vspace{38pt}
\end{minipage}%
\begin{minipage}[b]{0.79\textwidth}
  \begin{align}
  \label{eqn:susyTmult}
    \delta T^{\mu\nu} =&\ 
    \frac{1}{2} \varepsilon \gamma^{\rho (\mu} \partial_\rho J^{\nu)}\,,\nonumber\\
    \delta J^\mu =&\ 
    2 \varepsilon \gamma_{\nu} T^{\mu\nu}
    + \frac{2 c_2}{5 c_3} \left( 6 \eta^{\rho \mu}
    \left( \gamma^{\nu \sigma \lambda} + 3 \eta^{\sigma \nu} \gamma^\lambda
    \right)
    - \eta^{\mu\nu} \gamma^{\rho \sigma \lambda} \right)
    \check{\gamma}_I 
    \partial_\nu H_{\rho \sigma \lambda}^I\nonumber\\
    &+\frac{1}{10} \varepsilon
    \left( \gamma^{\mu\nu\rho} -4 \eta^{\mu \rho} \gamma^{\nu} \right)
    \check{\gamma}^{IJ}
    \partial_{\nu} j_{\rho IJ}\,,\nonumber\\
    \delta j^\mu_{IJ} =&\ 
    {-\frac{1}{2}} \varepsilon \check{\gamma}_{IJ} J^\mu
    + \frac{1}{5 c_3} \varepsilon \gamma^{\mu\nu}
    \partial_\nu \check{\gamma}_{[I} \chi_{J]}\,,\nonumber\\
    \delta H^I_{\mu\nu\rho} =&\ 
    \frac{c_3}{8 c_2} \varepsilon \check{\gamma}^I \gamma_{[\mu\nu} J_{\rho]}
    + \frac{1}{120 c_2} \varepsilon \gamma_\sigma \bar{\gamma}_{\mu\nu\rho}
    \partial^\sigma \chi^I\,,\nonumber\\
    \delta \chi^I =&\ 
    c_2 \varepsilon \gamma^{\mu\nu\rho}
    \left( \check{\gamma}^{IJ} + 4 \delta^{IJ} \right)
    H^J_{\mu\nu\rho} 
    + c_3 \varepsilon \gamma_\mu
    \left( \check{\gamma}^{IJK} + 3 \delta^{IJ} \check{\gamma}^{K} \right)
    j^{\mu}_{JK} \nonumber\\
    &+ \frac{1}{c_1} \varepsilon \gamma^\mu \check{\gamma}^{J}
    \partial_\mu O^{IJ}\,,\nonumber\\
    \delta O^{IJ} =&\ 
    c_1 \varepsilon \check{\gamma}^{(I} \chi^{J)}\,.
    \\[-5mm]\nonumber
\end{align}
\end{minipage}
\\
There are still some arbitrary constants $c_i$ that remain unfixed and can be
absorbed into the normalisations of $O, \chi$ and $H$. On the other hand,
the normalisation of the conserved currents must match that of the algebra, so
these operators cannot be rescaled. This can be seen by checking that the
variation of the currents reproduces the corresponding commutator
in~\eqref{eqn:ospfermionrep}. For example, the variation of $j^\mu$ computed
using~\eqref{eqn:susyTmult} is
\begin{align}
  \int \aQ_{\alpha \check{\alpha}} j^0_{IJ} \diff^5 x =
  -\frac{1}{2} \int \left( \check{\gamma}_{IJ} J^0 \right)_{\alpha \check{\alpha}} 
  \diff^5 x =
  -\frac{1}{2} \left( \check{\gamma}_{IJ} \aQ \right)_{\alpha \check{\alpha}}\,,
\end{align}
which is indeed the correct normalisation for the commutator $\left[ \aQ_{\alpha
\check{\alpha}}, \aR_{IJ} \right]$ of~\eqref{eqn:ospfermionrep}.

\subsection{Defect without insertions}
\label{sec:1PTstress}

Among the operators of the stress tensor multiplet, some can acquire an
expectation value in the presence of $V$. For the stress tensor, this happens
when $h_T \neq 0$ in~\eqref{eqn:stresstensor1pt}, and we can similarly constrain
the 1-point functions of the other operators. This computation is done
explicitly in Appendix~\ref{app:ConformalWardIdentities} and the only
nonvanishing correlators are
\bal
\label{eqn:StressTensor1Point}
  \vev{T^{ab} V } &= \frac{h_T \eta^{ab}}{\pi^3 x^6}\,, \qquad&
  \vev{T^{mn} V } &= -\frac{h_T}{\pi^3 x^6} \left(\delta^{mn} 
  - 2 \frac{x^{m} x^{n}}{x^2}\right)\,, \\
  \vev{H_{01m}^5 V } &= \frac{h_H x_{m}}{\pi^3 x^6}\,, \qquad&
  \vev{H_{lmn}^5 V } &= -\frac{h_H \varepsilon_{lmnp} x^{p}}{\pi^3 x^6}\,,\\
  \vev{O^{55} V } &= \frac{h_O}{\pi^3 x^4}\,, \qquad&
  \vev{O^{ij} V } &= -\frac{h_O \delta^{ij}}{4 \pi^3 x^4}\,,
\eal
where $h_O, h_H$, and $h_T$ are as yet undetermined constants. They are however
related by the supersymmetry transformations~\eqref{eqn:susyTmult} derived
above.  Specifically, consider the Ward identities associated with the preserved
supersymmetries $\aQ^+ = \Pi_+ \aQ$ (with the projector $\Pi_+$ defined
in~\eqref{eqn:ospprojector})
\begin{align}
\label{eqn:1PointSUSYWard}
\begin{split}
  0 &= \vev{\aQ^+_{\alpha \check\alpha}( \chi_{\beta \check\beta}^5 V)}
  = -4 \left( 12 c_2 h_H + \frac{h_O}{c_1} \right)
  \frac{\left[ \Pi_+ \gamma_m x^m \check\gamma^5 \right]_{\alpha \check\alpha
  \beta \check\beta}}{\pi^3 x^6}\,, \\
  0 &= \vev{\aQ^+_{\alpha \check\alpha} (J^a_{\beta \check\beta} V) } = 
  2 \left( h_T + \frac{36 c_2}{5 c_3} h_H \right)
  \frac{\left[ \Pi_+ \gamma^a \right]_{\alpha \check\alpha \beta \check\beta}}
  {\pi^3 x^6}\,.
\end{split}
\end{align}
These equations fix
\begin{align}
  h_O =  -12 c_1 c_2 h_H =  \frac{5}{3} c_1 c_3 h_T,
  \label{eqn:StressTensor1PointSUSY}
\end{align}
and the correlators in~\eqref{eqn:StressTensor1Point} are fixed up to a single 
constant $h_T$.

\subsection{Defect with an insertion}

We are now in a position to derive the result~\eqref{eqn:result} by relating 
$\vev{O^{i5}V[\bO^j]}$ to $\vev{T^{am} V[\bD^n]}$ using superconformal
Ward identities. There are two Ward identities to consider, $\vev{\aQ^+\chi
V[\bO]} = 0$ and $\vev{\aQ^+J V[\bD]} = 0$, but one can check that they yield the
same constraint, so we present only the first one.

The correlators we need are derived in
Appendix~\ref{app:ConformalWardIdentities} by using the constraints of conformal
symmetry. Importantly, the correlators $\vev{O V[\bO]}$, $\vev{\chi V[\bQ]}$ and
$\vev{H V[\bO]}$ are related to $h_T$ by integrated relations
like~\eqref{eqn:ExIntegrated}, while $\vev{j V[\bO]}$ is related to $C_\bD$
by~\eqref{eqn:ExDivergence}, as we show below. They are
\begin{align}
    \nonumber \vev{O^{i5} V[\bO^j]} &= \frac{C_{O \bO}\delta^{ij}}{x^2 (\sigma^2 +
    x^2)^2}\,,\qquad &
    \vev{\chi^5_{\alpha \check\alpha} V[\bQ_{\beta \check\beta}]} &= 
    \frac{C_{\chi \mathbb{Q}} \left[ \check{\gamma}^5 \left( \gamma_a \sigma^a
    + \gamma_{m} x^{m} \right)  \Pi_- c \Omega \right]
    _{\alpha \beta \check{\alpha} \check{\beta}}}{x^2 \left( x^2 + \sigma^2
    \right)^3}\,,\\
    \label{eqn:StressTensor2Point}
    \vev{j^{i5}_a V[\bO^j]} &= \frac{C_{j \bO} \delta^{ij}\sigma_a}{x^2
      (\sigma^2 + x^2)^{3}}\,, &
    \vev{j^{i5}_m V[\bO^j]} &= \frac{C_{j \bO} \delta^{ij} (x^2 - \sigma^2)
    x_m}{2 x^4 (\sigma^2 + x^2)^{3}}\,, \\
    \nonumber \vev{H^i_{01m} V[\bO^j]} &= \frac{C_{H \bO}\delta^{ij} x_m}{x^2 (\sigma^2 +
    x^2)^3}\,, &
    \vev{H^i_{lmn} V[\bO^j]} &= \frac{C_{H \bO}\delta^{ij} 
    \varepsilon_{lmnp}x^p}{x^2 (\sigma^2 + x^2)^3}\,.
\end{align}
Explicitly, the Ward identity is
\begin{align}
\label{eqn:SUSYWardIdentity-ChiO}
  \begin{aligned}
    0 &= \vev{\aQ^+_{\alpha \check{\alpha}} \left( \chi^5_{\beta \check{\beta}} V[\bO^{i}] \right)}\\
    &= 6c_2 \left[ \Pi_+ \gamma^{01m} ({\check\gamma}\indices{^5_J}
      + 4 \delta^5_J) \right]_{\alpha \check\alpha \beta \check\beta} \vev{H^J_{01m} V[\bO^i]}
      + 6c_2 \left[ \Pi_+ \gamma^{lmn} ({\check\gamma}\indices{^5_J} + 4 \delta^5_J) \right]
      _{\alpha \check\alpha \beta \check\beta} \vev{H^J_{lmn} V[\bO^i]} \\
      &\quad{}
      + 3c_3 \left[ \Pi_+\gamma^\mu {\check\gamma}_j \right]_{\alpha \check\alpha \beta \check\beta}
      \vev{j^{5j}_\mu V[\bO^i]}
      + \frac{1}{c_1} \left[ \Pi_+ \gamma^\mu \check\gamma_J \right]
      _{\alpha \check\alpha \beta \check\beta} \partial_\mu \vev{O^{5 J} V[\bO^i]}\\
      &\quad{}
      + \frac{1}{2}({\check\gamma}^{i5})\indices{_{\check\alpha}^{\check\gamma}} 
    \vev{\chi^5_{\beta \check\beta} V[\bQ_{\alpha \check\gamma}]}\,.
  \end{aligned}
\end{align}
Plugging in the explicit forms of these
correlators~\eqref{eqn:StressTensor2Point}, and demanding that the terms
proportional to $\gamma_a \sigma^a$ vanish, we obtain a linear relation
\begin{align}
  0 = 3 c_3 C_{j \bO} + \frac{4}{c_1} C_{O \bO} - C_{\chi \bQ}\,.
  \label{eqn:swardidentity}
\end{align}
The terms proportional $\gamma_m x^m$ give the same constraint.

Next, recall that $\bO$ and $\bQ$ respectively encode the action of a broken
infinitesimal R-symmetry or supersymmetry variation. Therefore we can relate
\beq
0 = \vev{\aR_{j 5} (O^{i 5}(x) V)}= \delta^{ij} \vev{O^{55}(x) V} - \vev{O^{ij}(x) V}
  + \int \diff^2 \sigma \vev{O^{i5}(0,x) V[\bO^j](\sigma)}\,.
\eeq
Using \eqref{eqn:StressTensor1Point} and \eqref{eqn:StressTensor2Point}, we obtain
\begin{align}
  C_{O \bO} = - \frac{5}{4 \pi^4} h_O = -\frac{25c_1 c_3}{12 \pi^4} h_T\,.
\end{align}
A slightly more involved but entirely analogous calculation yields 
\begin{align}
  C_{\chi \bQ} = -\frac{5 \cdot 8}{3 \pi^4} h_T\,, \qquad
  C_{H \bO} = \frac{5 c_3}{36 c_2 \pi^4} h_T\,.
\end{align}
Finally, $C_{j\bO}$ is related to the normalisation of the displacement operator
multiplet by~\eqref{eqn:broken}
\begin{align}
\partial_\mu \vev{j^{\mu i 5}(\sigma, x) V[\bO^j(0)]} = \vev{V[\bO^i(0)
\bO^j(\sigma)]} \delta^{(4)}(x)\,.
\end{align}
Plugging the correlator of $j^{\mu i5}$ and $\bO^j$ into the right hand side and integrating 
against a test function allows us to fix 
\begin{align}
\label{eqn:Conservation}
  C_{j\bO} = -\frac{1}{\pi^4} C_{\bO} = -\frac{1}{16 \pi^4} C_\bD\,.
\end{align}
Combining the above results into~\eqref{eqn:swardidentity}, we obtain
\begin{align}
  \frac{c_3}{\pi^4} \left( 3 C_\bO - 5 h_T \right) = 0
  \qquad \Longrightarrow \qquad
  h_T = \frac{3 C_\bO}{5} = \frac{3 C_\bD}{80}\,,
  \label{eqn:swardidentityresult}
\end{align}
which proves~\eqref{eqn:result}.

\section{Relation to anomaly coefficients}
\label{sec:anomaly}

In this section we explore the consequences of the relation between the
coefficients $C_\bD$ and $h_T$~\eqref{eqn:result} for physical observables.  
These pieces of dCFT data appear in the Weyl anomaly of surface operators 
as defined in~\eqref{eqn:anomaly}, and as we show below the relations~\eqref{eqn:cdco}
and~\eqref{eqn:swardidentityresult} relate the anomaly coefficients
as~\eqref{eqn:anomalyresult}.

The relation between correlators and anomaly coefficients is not specific to 
2d defects in the $\mathcal{N} = (2,0)$ theory, but applies for any 
surface operator in a CFT. The anomaly coefficient $b_1$ was first shown to be 
related to $C_\bD$ in~\cite{bianchi:2015liz}, while the relation between $b_2$ 
and $h_T$ was obtained in~\cite{Lewkowycz:2014jia,bianchi:2015liz}.
Here we review their derivation and apply it to surface operators in the (2,0)
theory to prove $c = -b_1/2$, $b_1 = -b_2$.

In a slightly different direction, the anomaly coefficients also feature notably
in entanglement entropy in 4d~\cite{solodukhin:2008dh} and were discussed in the
entanglement entropy literature, see~\cite{bianchi:2015liz} and references
therein.

\subsection{Displacement operator}

In order to isolate the contribution of $C_\bD$ to the anomaly coefficients, 
we separately switch on each of the terms in~\eqref{eqn:anomaly}. 
Since the displacement operator generates geometric deformations, one expects that
inserting sufficiently many $\bD^m$ into the planar surface operator $V$ leads to 
a logarithmic divergence in the expectation value, signalling a conformal anomaly 
associated to the curvature of the surface. Similarly, inserting $\bO^i$ to sufficient 
order will allow us to access the anomaly coefficient $c$ associated with deformations 
in R-symmetry space.

To make this relation precise, we formally write deformations of the 1/2
BPS plane in terms of operator insertions
\begin{align}
  V_{\xi,\omega} = \exp{\left[ \int \diff^2 \sigma \xi_m(\sigma)
    P^m
  + \omega_{i}(\sigma) R^{i5} \right]} V.
  \label{eqn:WSdeforVEV}
\end{align}
Here $P^m = \int \diff^4 x \partial_\mu T^{\mu m}$ generates translations
transverse to the defect, while R-symmetry rotations are generated by 
$R^{i5} = \int \diff^4 x \partial_\mu
j^{\mu i5}$. For constant parameters $\xi, 
\omega$, the currents can be freely integrated and we recover the standard 
action of the charges $\aP^m$ and $\aR^{i5}$.

Equation~\eqref{eqn:WSdeforVEV} is generally a complicated expression
involving contact terms like~\eqref{eqn:dispop}, but also contact terms from
$P^m$ acting on defect operators and possibly other operators from the OPE.
We can calculate its expectation value to quadratic order by expanding the
exponential and noting that the 1-point functions of defect operators vanish:
\begin{equation}
  \log{\vev{V_{\xi,\omega}}} - \log{\vev{V}} =
  \frac{1}{2} \int_{{\mathbb R}^2 \times {\mathbb R}^2}
  \Big( \vev{V[\bD_{m} \bD_{n}]} \xi^{m} \xi^{n}
  + \vev{V[\bO_{i} \bO_{j}]} \omega^{i} \omega^{j} \Big)
  \label{eqn:vevdisp}
 \diff^2\sigma\,\diff^2\sigma'+\text{cubic.}
\end{equation}
We can discard $\log{\vev{V}}$ since for the 1/2-BPS plane in a flat background, all anomaly terms 
vanish separately. Since the anomaly is quadratic in $\xi$ and $\omega$, it is 
related to the two point functions written here and we can safely ignore the higher 
order terms in the expansion.

To extract the anomaly coefficients, we study the UV divergence of the integrals 
in~\eqref{eqn:vevdisp}. The relevant correlators are found
in~\eqref{eqn:disp2pts} and~\eqref{eqn:DISP2pts}.
Fixing $\sigma$, the $\sigma'$ integral can be evaluated explicitly by 
Taylor expanding $\xi^{m}(\sigma')$ and $\omega^{i}(\sigma')$ around $\sigma$. 
Starting with the second integrand and using $\tau=\sigma'-\sigma$,
\begin{equation}
\begin{aligned}
  \frac{1}{2} \int_{\mathbb R^2} &\vev{V[\bO_{i}(\sigma) \bO_{j}(\sigma')]} 
  \omega^{i}(\sigma)
  \omega^{j}(\sigma')\,\diff^2\sigma'
  \\&= \frac{C_\bO}{2 \pi^2} \int_{\mathbb R^2} \frac{\delta_{i j }}{\norm{\tau}^4}
  \omega^{i}(\sigma)\left[\omega^{j}(\sigma)+\tau^a\partial_a\omega^{j}(\sigma)
  +\frac{1}{2}\tau^a\tau^b\partial_a\partial_b\omega^{j}(\sigma)+\ho[3]{\tau}\right]\diff^2\tau.
\end{aligned}
\end{equation}
While this integral leads to power law singularities as well, a logarithmic 
divergence arises only from the term quadratic in $\tau$. We adopt 
polar coordinates $\tau^a = \tau e^a$ where $e^a$ are orthonormal vectors parametrised 
by an angle $\varphi$. Using the identities
\begin{align}
  \int \diff \varphi \, e^{a} e^{b} = \pi  \eta^{ab}, \qquad
  \int \diff \varphi \, e^{a} e^{b} e^c e^d = 
  \frac{\pi }{4} (\eta^{ab} \eta^{cd} + \eta^{ac} \eta^{bd} +
  \eta^{ad} \eta^{bc}) \, , 
  \label{eqn:angularidentities}
\end{align}
and dropping all but the logarithmic divergence, we obtain
\begin{equation}
\frac{C_\bO}{4\pi^2} \pi \eta^{ab} \int\limits_{\epsilon} \frac{\tau^3 \diff \tau}{\tau^4}
  \omega^{i}(\sigma) \partial_a\partial_b \omega^{i}(\sigma) 
= \frac{C_\bO}{4 \pi} (\partial\omega)^2 \log{\epsilon}.
\label{eqn:coIntermediate}
\end{equation}
To leading order, the R-symmetry transformation in~\eqref{eqn:WSdeforVEV}
takes the 1/2-BPS plane to a surface operator with $\partial_a n^i(\sigma) = \partial_a \omega^i$, 
so we can read the anomaly coefficient as
\begin{align}
  c=C_\bO.
  \label{eqn:cco}
\end{align}

The logarithmic divergence of the first integrand in~\eqref{eqn:vevdisp} can be evaluated in 
a similar way, and arises only from the fourth order in the Taylor expansion of
$\xi^{n}$
\begin{equation}
\begin{aligned}
&  \frac{1}{2} \int_{\mathbb R^2} \vev{V[\bD_{m} \bD_{n}]}
\xi^{m} \xi^{n} \diff^2\sigma'
\\&\hskip2cm
= \frac{C_\bD}{2 \pi^2} \int_{\mathbb R^2} \frac{\delta_{mn}}{\norm{\tau}^6}
\xi^{m}(\sigma)\left[\cdots+\frac{1}{24}\tau^a\tau^b\tau^c\tau^d\partial_a\partial_b\partial_c\partial_d
  \xi^{n}(\sigma)+\ho[5]{\tau}\right]
\diff^2\tau\,.
\end{aligned}
\end{equation}
Performing the angular integral with~\eqref{eqn:angularidentities} leads to
\beq
  \frac{C_\bD}{48 \pi^2}\frac{3\pi^2}{4} \int\limits_\epsilon
  \frac{\tau^5 \diff \tau}{\tau^6} \xi_{m}(\sigma) (\partial^2)^2\xi^{m}(\sigma)
  =
  -\frac{C_\bD}{64 \pi}  
    \partial^a\partial^b \xi_{m}(\sigma) \partial_a\partial_b \xi^{m}(\sigma)\log{\epsilon}\,.
\eeq
This is the trace of the second fundamental form squared of the deformed surface
(see~\eqref{eqn:IIFundForm}), which can be rewritten 
using the Gauss-Codazzi equation~\eqref{eqn:GaussCodazzi} as
\beq
  \partial^a\partial^b \xi_{m} \partial_a\partial_b \xi^{m} = \IIFundForm^2 =
  2 \tr{\tilde{\IIFundForm}^2} + \RicciScalar^{\Sigma} - \tr{W}\,.
  \label{eqn:DispIIFundForm}
\eeq
Since we are on flat space, the Weyl tensor vanishes.
The volume form for the deformed surface gets corrected, but to
leading order in $\xi$ does not affect the calculation. Therefore the 
contribution of this term to the anomaly density is
\beq
  -\frac{C_\bD}{64 \pi}
  \int_{\Sigma} \left( 2 \tr{\tilde{\IIFundForm}^2} + \RicciScalar^{\Sigma} \right) \vol_{\Sigma}
  \log{\epsilon} \,.
\eeq
Note that the integral of $\RicciScalar^{\Sigma}$ vanishes for small
deformations of the plane.  It therefore does not contribute to the anomaly, and we find
\begin{align}
  b_1=-C_\bD/8\,.
  \label{eqn:b1cd}
\end{align}
Using~\eqref{eqn:cdco} along with~\eqref{eqn:cco} and~\eqref{eqn:b1cd} we find a
relation for the anomaly coefficients
\begin{align}
  c = -b_1/2\,.
  \label{eqn:cb1}
\end{align}

\subsection{Stress tensor}

The relation between $b_2$ and $h_T$ is derived in a similar fashion, but instead of 
deforming the surface itself, we can relate the insertion of a stress tensor to a change 
in the background geometry.\footnote{%
  In the same way one can show that the bulk anomaly coefficients are related to
  the 2- and 3-point functions of the stress tensor~\cite{Bastianelli:2000hi}.
}
The expectation value of the planar 
surface operator now receives a contribution from the metric variation:
\begin{align}
  \vev{V}_{\eta + \delta g} = \vev{V}_\eta
  - \frac{1}{2} \int_{\mathbb{R}^2 \times \mathbb{R}^4}
  \delta g_{\mu\nu}(\sigma,x) \vev{T^{\mu\nu} V}_{\eta}
  \diff^2 \sigma \diff^4 x + \ho[2]{\delta g}\,.
  \label{eqn:VTexpansion}
\end{align}
In this equation, the subscript $\vev{\bullet}_g$ means the expectation value is
calculated on a curved background metric $g$. 

Since the insertion of a stress tensor sources a metric perturbation of linear
order $\delta g$, we can only reproduce the anomaly to that order, which,
expanding~\eqref{eqn:anomaly}, is
\begin{align}
  \mathcal{A}|_{\delta g} = \frac{1}{4 \pi} \left[ -\frac{b_2}{10} \left(
    \partial_{p}^2 \delta^{mn} - \partial^{m} \partial^{n} \right) \delta g_{mn}
  + \frac{3b_2}{20} \eta^{ab} \partial_{p}^2 \delta g_{ab} + \partial_a (\dots)
\right].
\label{eqn:anomalydg}
\end{align}
These two terms are respectively associated to $\vev{T^{mn} V}$ and
$\vev{T^{ab} V}$ in~\eqref{eqn:VTexpansion}, and the total derivative drops out
of the integral over the plane.

Using~\eqref{eqn:StressTensor1Point}, we can evaluate the first term
of~\eqref{eqn:VTexpansion}. The logarithmic divergence arises as
\begin{align}
  \int_{\mathbb{R}^4} \delta g_{mn} \vev{T^{mn} V} \diff^4 x
  &= -\frac{h_T}{\pi^3}
  \int_{\mathbb{R}^4} \diff^4 x \delta g_{mn}(\sigma,x) \frac{\delta^{mn} - 2
  x^{m} x^{n}/x^2}{x^6}\\
  &= -\frac{h_T}{\pi^3}
  \int_{\mathbb{R}^4} \frac{\diff^4 x}{x^6} \left( \dots +
  \frac{1}{2} \left. \partial_{p q} \delta g_{mn} \right|_{x=0} x^{p} x^{q} + \dots \right)
  \left(\delta^{mn} - 2 \frac{x^{m} x^{n}}{x^2}\right)\,.
  \nonumber
\end{align}
In the second step we expanded $\delta g(x)$ in a Taylor series and
dropped powers of $x$ not contributing to the anomaly.  We again switch to 
spherical coordinates $x^m = r e^m$ and take note of the 4d analogue 
of~\eqref{eqn:angularidentities}
\begin{align}
  \int \vol_{S^3} e^{m} e^{n} = \frac{\pi^2}{2} \delta^{mn}\,, \qquad
  \int \vol_{S^3} e^{m} e^{n} e^{p} e^{q} = \frac{\pi^2 }{12} \left(
    \delta^{mn} \delta^{pq} + \delta^{mp} \delta^{nq} +
    \delta^{mq} \delta^{np} \right)\,.
\end{align}
The integral then becomes
\begin{equation}
    -\frac{2 \pi^2 h_T}{2\pi^3} \frac{2}{3} 
    \int_{\epsilon} \frac{\diff r}{r} \left. \partial_{p q} \delta g_{mn} \right|_{x=0}
    \left(
      \delta^{mn} \delta^{pq} - \delta^{mp} \delta^{nq} \right)
    = \frac{1}{4 \pi} \log{\epsilon} \left[ \frac{2 h_T}{3}
      \left( \partial^2_{p} \delta^{mn} - \partial^{m} \partial^{n} \right)
    \delta g_{mn} \right]_{x=0}\,.
\end{equation}
Comparing against~\eqref{eqn:anomalydg}, we identify
\begin{align}
  h_T = \frac{3 b_2}{10}\,.
  \label{eqn:CTb2relation}
\end{align}
The calculation for $\vev{T^{ab} V}$ is similar and gives the same result.

With expressions for $b_1, b_2, c$ in terms of $C_\bD$ and $h_T$ in hand, we can 
finally translate the result of the previous section~\eqref{eqn:swardidentityresult} 
into a constraint on the anomaly coefficients, and find
\begin{align}
  b_2 = -b_1\,,
  \label{eqn:b2b1}
\end{align}
as claimed.

A direct consequence of this relation (together with~\eqref{eqn:cb1}) is that
one only needs to calculate two nontrivial surface operators to calculate all the
independent anomaly coefficients, for instance the sphere and cylinder.

\section{Defect operator expansion}
\label{sec:doe}

A useful tool in dCFT is the defect operator expansion (dOE),
also known as the bulk-defect operator product
expansion~\cite{Diehl:1981jgg,Cardy:1991tv} (see~\cite{Liendo:2012hy} for a
recent review of some dCFT techniques, including the dOE, in the context of the
CFT bootstrap program).  This is a convergent expansion representing bulk operators
in terms of insertions of defect operators
\begin{align}
  \cO_i (\sigma, x)V = \sum_{k} \frac{C^V_{ik} (x,\partial_\sigma)}
  {x^{\Delta_i - \hat{\Delta}_k}} V[\hat\cO_k(\sigma)]\,,
  \label{eqn:doe}
\end{align}
where the sum is over defect primaries. The differential operators $C^V_{ik}(x,
\partial_\sigma)$ are fixed by conformal symmetry. Their exact form can be obtained 
from the corresponding bulk-defect 2-point function of $\cO_i$ and $\hat\cO_k$ by equating
\begin{align}
\label{eqn:doeVev}
  \begin{split}
    \vev{\cO_i(\sigma, x) V[\hat\cO_k(0)]} &= \sum_j
    \frac{C^V_{ij} (x,\partial_\sigma)}{x^{\Delta_i - \hat{\Delta}_j}}
  \vev{V[\hat\cO_j(\sigma) \hat\cO_k(0)]}
  = \frac{1}{x^{\Delta_i - \hat{\Delta}_k}} C^V_{ik} (x,\partial_\sigma)
  \frac{C_{\hat\cO_k}}{\sigma^{2\hat\Delta_k}}\, ,
  \end{split}
\end{align}
where we denote by $C_{\hat\cO_k}$ the numerator of the 2-point function of $\hat{\cO}_k$.
Explicit expressions for $C_{ik}^V$ can be found in~\cite{McAvity:1995zd,Billo:2016cpy}, 
but are not needed in this paper.

The list of defect primaries appearing on the right-hand side
of~\eqref{eqn:doe} can include the defect operators of
Section~\ref{sec:BulkDefectCorrelators} (namely the defect identity
and the displacement operator multiplet), but it certainly includes
more defect operators. This can be viewed as a consequence of the associativity
of the OPE: since~\eqref{eqn:doe} maps bulk operators to defect operators and is
valid in any correlator, all the CFT data of the bulk operators must be encoded,
in some way, in the OPE of defect operators. Hence there must be at least as
many defect degrees of freedom as bulk degrees of freedom.

Here we initiate the study of these other defect operators. We first classify the
unitary multiplets of defect operators in Sections~\ref{sec:dops}
and~\ref{sec:unitarity}. This allows us to find the decomposition of the stress
tensor multiplet in multiplets of the preserved algebra, see
Figures~\ref{fig:stmult} and~\ref{fig:stmult2}.

After this detour into representation theory, we write the leading terms in 
the dOE for some operators and discuss the appearance of a new marginal
operator. We finally comment on constraints imposed by supersymmetry and show
how the dOE sheds light on the derivation of Section~\ref{sec:BulkDefectCorrelators}.

\subsection{Representations of
  \texorpdfstring{$\osp(4^*|2) \oplus \osp(4^*|2)$}{osp(4|2)+osp(4|2)}}
\label{sec:dops}

Defect operators sit in multiplets of the algebra preserved by the defect. For the
1/2-BPS plane $V$, the preserved algebra consists of 2 copies of $\osp(4^*|2)$, so we
are interested in constructing representations of $\osp(4^*|2) \oplus \osp(4^*|2)$. The
formulation of the algebra as a 2d superconformal algebra is reviewed in the
appendix~\ref{sec:ospsubalg}, along with its embedding inside the bulk
algebra $\osp(8^*|4)$.

As usual, we can label primaries by their representation under the bosonic
subalgebra, which here is
\begin{align}
  \left[ \sl(2) \oplus \su(2)_\perp \oplus \su(2)_R \right] \oplus
  \left[ \sl(2) \oplus \su(2)_\perp \oplus \su(2)_R \right].
  \label{eqn:boosnicsubalgebraosp}
\end{align}
The corresponding labels are $[r_1, r_2]_{h} [\bar{r}_1, \bar{r}_2]_{\bar{h}}$,
with $r_1$ and $r_2$ the Dynkin labels for $\su(2)_\perp$ and $\su(2)_R$, and $h$ the
conformal twist and labels representations of $\sl(2)$.  The labels $\bar{r}_1$,
$\bar{r}_2$ and $\bar{h}$ are similar, but for the second subalgebra.
We note that while~\eqref{eqn:boosnicsubalgebraosp} is equivalent to
$\so(2,2)_\parallel \oplus \sof(4)_\perp \oplus \so(4)_R$, the factorisation
above in terms of 2 algebras is dictated by supersymmetry,
see~\ref{sec:ospsubalg} for more details.  The joint representation has conformal 
dimension $\hat{\Delta} = h +\bar{h}$ and spin $s = h - \bar{h}$.

The simplest nontrivial example of a multiplet of $\osp(4^*|2) \oplus \osp(4^*|2)$ is the familiar
displacement multiplet of section~\ref{sec:disp}. Unlike our previous treatment however,
here we label operators according to~\eqref{eqn:boosnicsubalgebraosp}.
In order to match that decomposition, we
can express the superprimary $\bO^i \sim
(\check{\gamma}^i)^{\alpha_2 \dot{\alpha}_2} \bO_{\alpha_2 \dot{\alpha}_2}$ in
spinor indices. In this notation, the indices $\alpha = 1,2$ are all $\su(2)$ indices. We
use $\alpha_1, \beta_1, \dots$ for $\su(2)_\perp$ and $\alpha_2, \beta_2, \dots$
for $\su(2)_R$; similarly for the second set of $\su(2)$'s, but with dotted indices.

The values of $h$ and $\bar{h}$ can also be read from~\eqref{eqn:broken}, they
are $h = \bar{h} = 1$ ($\bO$ is a scalar of dimension 2). The representation of
$\bO$ is therefore $[0,1]_1 [0,1]_1$.
Acting with $\aQ$ and $\bar{\aQ}$ (which transform respectively as
$[1,1]_{1/2}[0,0]_0$ and $[0,0]_0 [1,1]_{1/2}$), one can build the full
multiplet:

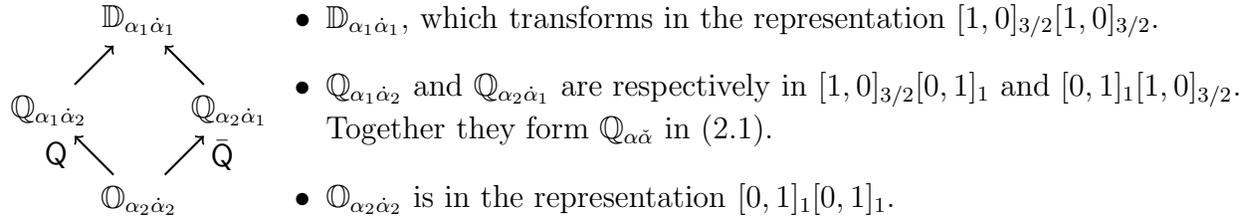
\newcommand{\tikzmark}[1]{\tikz[overlay,remember picture] \node (#1) {};}
\noindent
\begin{minipage}[b]{0.2\textwidth}
  \begin{tikzpicture}[scale=2.4,baseline=(current bounding box.south)]
    \node (nh) at (0,2) {$\bD_{\alpha_1 \dot{\alpha}_1}$};
    \node (nc1) at (-0.5,1.5) {$\bQ_{\alpha_1 \dot{\alpha}_2}$};
    \node (nc2) at (0.5,1.5) {$\bQ_{\alpha_2 \dot{\alpha}_1}$};
    \node (no) at (0,1) {$\bO_{\alpha_2 \dot{\alpha}_2}$};
    \draw[thick,->] (no) edge node[left,anchor=east,xshift=-2mm] {$\aQ$}
    (nc1) (nc2) edge (nh);
    \draw[thick,->] (no) edge node[right,anchor=west,xshift=2mm] {$\bar{\aQ}$}
    (nc2) (nc1) edge (nh);
  \end{tikzpicture}
  \vspace{-7pt}
\end{minipage}%
\begin{minipage}[b]{0.8\textwidth}\raggedright\vspace{8pt}
\begin{itemize}
  \item $\bD_{\alpha_1 \dot{\alpha}_1}$, which transforms in the representation
    $[1,0]_{3/2} [1,0]_{3/2}$.
  \item $\bQ_{\alpha_1 \dot{\alpha}_2}$ and $\bQ_{\alpha_2 \dot{\alpha}_1}$ are
    respectively in $[1,0]_{3/2} [0,1]_1$ and $[0,1]_1 [1,0]_{3/2}$. Together
    they form $\bQ_{\alpha \check{\alpha}}$ in~\eqref{eqn:broken}.
  \item $\bO_{\alpha_2 \dot{\alpha}_2}$ is in the representation $[0,1]_1 [0,1]_1$.
\end{itemize}
\vspace{\parskip}
\end{minipage}

The structure of the multiplet as a product of two representations of
$\osp(4^*|2)$ is apparent in the diagram above.
Under the action of $\aQ$, the operators transform as two multiplets of
$\osp(4^*|2)$, for instance the lower diagonal is
\begin{align}
  \aQ_{\alpha_1 \alpha_2} \bO_{\beta_2 \dot{\beta}_2}
  = c \epsilon_{\alpha_2 \beta_2} \bQ_{\alpha_1 \dot{\beta}_2}\,, \qquad
  \aQ_{\alpha_1 \alpha_2} \bQ_{\beta_1 \dot{\beta}_2}
  = i c^{-1} \epsilon_{\alpha_1 \beta_1} \partial \bO_{\alpha_2
    \dot{\beta}_2}\,,
\label{eqn:dispSusyTrafos}
\end{align}
which is easily obtained from an ansatz as in Section~\ref{sec:stress} (the
constant $c$ is arbitrary). This is the simplest representation of $\osp(4^*|2)$
and it contains the weights $[0,1]_1$ and $[1,0]_{3/2}$. Because it is
ubiquitous, it is convenient to introduce some notation here and denote it 
$B[0,1]$, in anticipation of the results of Section~\ref{sec:unitarity}.

\subsection{Unitary multiplets of \texorpdfstring{$\osp(4^*|2)$}{osp(4|2)}}
\label{sec:unitarity}

Since the algebra preserved by the defect factorises, we now turn our focus to
general multiplets of a single copy of $\osp(4^*|2)$. 
Importantly, we can classify allowed multiplets by working out the constraints
imposed by unitarity.\footnote{%
  The same analysis was also done in~\cite{Agmon:2020pde}, which appeared as
  this paper was finalised.
}
This follows the method described in~\cite{minwalla:1997ka} used to classifiy
multiplets in superconformal theories for $d \ge 3$.

The idea is the following. In radial quantisation, any operator
$\cO$ defines a corresponding state $\ket{\cO}$.  While $\ket{\cO}$ has positive
norm (by assumption), there is no guarantee that the norm of all the other
states of the multiplet is also positive, as required by unitarity. Demanding
that negative norm states are absent from the multiplet leads to a lower bound
on the conformal dimension of the superprimary $h \ge h_A$. In particular, as we
show below, at $h = h_A$~\eqref{eqn:ha} some states become null, and the
corresponding multiplets are the short multiplets $A$. In addition, we find yet
shorter multiplets $B$ with superprimary of conformal dimension
$h_B$~\eqref{eqn:hb}.

Consider the state $\ket{\cO}$ of a superprimary operator in the representation
$[r_1,r_2]_h$. Unitarity constrains the states $\aQ\ket{\cO}$ to satisfy
\begin{align}
  \left\|\aQ\ket{\cO}\right\|^2
  =\bra{\cO}\{\aS,\aQ\}\ket{\cO}
  =  \bra{\cO} D_+ + \sigma^i \aT_{(1)}^i - 2 \sigma^j \aT_{(2)}^j \ket{\cO}
  \ge 0\,,
  \label{eqn:unitaritycons}
\end{align}
where we use $\aQ_{\alpha_1 \alpha_2}^\dagger = \aS^{\alpha_1 \alpha_2}$
and the anticommutator~\eqref{eqn:osp4susy}, written in terms of $\su(2)_\perp$
and $\su(2)_R$ generators $\aT^i_{(1,2)}$. We suppress the indices of $\aQ$ and $\ket{\cO}$, but
the constraint should hold for any choice of $\aQ$, $\ket{\cO}$, and linear
combinations thereof.

The matrix elements $\vev{s| \sigma^i \aT^i |s}$ are bounded by the eigenvalues of
$\sigma^i \aT^i$. Since $\sigma^i$ is the fundamental representation, the product
$\sigma^i \aT^i$ can be decomposed as $[1] \otimes [r] = [r-1] \oplus [r+1]$, for both
$r_1$ and $r_2$. The eigenvalues are expressed in terms of the quadratic Casimirs
$C_2(j) = j(j+2)/4$ (using e.g. equation (2.38) of~\cite{minwalla:1997ka}), so
that~\eqref{eqn:unitaritycons} takes the form
\begin{align}
  \label{eqn:unitarityboundGeneric}
  h \geq - \left( C_2(j_1) - C_2(1) - C_2(r_1) \right) + 2 \left( C_2 (j_2) -
  C_2(1) - C_2(r_2) \right),
\end{align}
with $j_1$ and $j_2$ taking any values in $r_1 \pm 1$ and $r_2 \pm 1$. This
assumes that both $r_1 > 0$ and $r_2 > 0$, otherwise the tensor product
decomposition is simply $[1] \otimes [0] = [1]$ and $j = 1$.

For $r_1 > 0$, we then find that the strongest bound on the scaling dimension
implied by~\eqref{eqn:unitarityboundGeneric} is 
\begin{align}
  h \geq h_A = 1 + \frac{r_1}{2} + r_2\,.
  \label{eqn:ha}
\end{align}
For $r_1=0$, we should instead take $j_1 = 1$ and we obtain
\begin{align}
  h \geq h_B = r_2, \qquad \text{if}\quad r_1 = 0\,.
  \label{eqn:hb}
\end{align}
If these bounds are saturated, a subset of states become null and may be
consistently removed from the multiplet.

While~\eqref{eqn:ha} and~\eqref{eqn:hb} are necessary conditions for unitarity, 
there could be, in principle, additional states whose norm becomes null (or
negative), imposing further restrictions on $h$. It would be tedious to perform
the above calculation for all states, but fortunately the conditions under which
a representation is reducible (but not necessarily unitary) are listed by Kac
in~\cite{kac1978representations} (see also~\cite{vanderjeugt:1985hq}). These match
precisely the values obtained for the 4 choices of $j_1$ and $j_2$
in~\eqref{eqn:unitarityboundGeneric}, which indicates that there are no further
constraints.

We therefore conclude that for multiplets satisfying $h \ge h_A$, with $h_A$ given
in~\eqref{eqn:ha}, there are no stronger constraints from requiring unitarity at higher 
levels. Generically, these are long multiplets, and they thus contain $2^4
(r_1+1)(r_2+1)$ operators. Multiplets saturating the bound $h = h_A$ have a null
state at level one, $\ket{[r_1-1,r_2+1]_{h+1/2}}$, and their dimension is
reduced. The special case $r_1 = 0$ still leads to a unitary multiplet, but in
this case the first null state is at level 2.

In the case $h_A > h \ge h_B$~\eqref{eqn:hb} however, since $h$ is below the unitarity
bound $h_A$, some states in the multiplet would have a negative norm unless $h =
h_B$ exactly: this is an isolated multiplet. It has a null state at level one,
$\ket{[1, r_2+1]_{h+1/2}}$.

These short multiplets $A$ and $B$ are important to our discussion. For  
example, the $B[0,1]$ multiplet of Section~\ref{sec:dops} contains only $2+2$
operators, so it is indeed a short multiplet. From the argumentation above, the
conformal dimension of its superprimary is thus fixed by unitarity to $h = h_B =
1$, in accordance with~\eqref{eqn:broken}.

The broader question of determining the content of all short
multiplets is interesting but lies beyond the scope of this work.
However, specific short multiplets play a role in
Section~\ref{sec:repsex}, and it is useful to know their
content explicitly. It is sufficient for our present purposes to construct some
representations heuristically by taking the tensor product decomposition of
known multiplets.  For instance, taking the product of two $B[0,1]$ multiplets,
the superprimary decomposes into 2 representations $[0,1] \otimes [0,1] = [0,0] 
\oplus [0,2]$, so the tensor product gives 2 multiplets, which we identify as
\begin{align}
  B[0,1] \otimes B[0,1] = A[0,0] \oplus B[0,2]\,.
\end{align}
The multiplet $A[0,0]$ contains the weights $[0,0]_1, [1,1]_{3/2}$ and
$[2,0]_2$, while the multiplet $B[0,2]$ contains $[0,2]_2, [1,1]_{5/2}$ and
$[0,0]_3$. Both of these representations appear as defect operators, see 
Figures~\ref{fig:stmult} and~\ref{fig:stmult2} below.

\subsection{The stress tensor dOE}
\label{sec:repsex}

Having gained some understanding of representations of the preserved algebra, we 
turn now to the main goal of this section: constructing the dOE~\eqref{eqn:doe}
for the bulk operators of our theory. We focus on operators of the stress tensor
multiplet (which should exist in any local quantum field theory), but the
same analysis could be applied to other multiplets.

A naive way of thinking about~\eqref{eqn:doe} is as branching rules for the
breaking of symmetry due to the presence of the defect. Indeed, it is natural to
decompose, for example, the bulk superprimary $O^{IJ}$ into representations of
the preserved R-symmetry $O^{55}, O^{i5}$ and $O^{ij}$, respectively the
representations
\begin{align}
  [0,0] [0,0]\,, \qquad
  [0,1] [0,1]\,, \qquad
  [0,2] [0,2]\,.
\end{align}
The dOE~\eqref{eqn:doe} is particularly simple for a trivial surface 
defect, where it is just the Taylor expansion of the bulk insertion:
\begin{align}
  O^{55}(x) I = I[O^{55}(0) + x^m \partial_m O^{55}(0) + \dots]\,,
\end{align}
While this expression merely amounts to a rewriting of the bulk degrees 
of freedom, the dOE becomes much more interesting if we consider a defect 
$V$ which interacts with the bulk nontrivially.

A first sign that the dOE for general $V$ contains additional terms is that the
bulk operators couple to the defect identity $\bf{1}_V$ and the displacement
multiplet (cf. for instance~\eqref{eqn:StressTensor1Point} and~\eqref{eqn:StressTensor2Point}). 
It is clear that these operators do not appear in the branching rules and encode 
additional interactions between bulk and defect degrees of freedom.

The second way in which the dOE is interesting is more subtle. The decomposition
of operators in terms of the preserved algebra can be performed, as above, for all
the operators in the stress tensor multiplet. The resulting representations can be
organised in the multiplets of Figures~\ref{fig:stmult} and~\ref{fig:stmult2}
and the displacement multiplet, leading to the branching rules under
the breaking of symmetry $\osp(8^*|4) \to \osp(4^*|2) \oplus \osp(4^*|2)$. 
The superprimaries of the multiplets in Figure~\ref{fig:stmult} are easily
identified as the defect counterparts of the operators $O^{55}$ and $O^{ij}$ by
their representation, and with a bit of work this correspondence between bulk and defect
operators can be also established for all the other operators.


\definecolor{pnodes}{rgb}{0.4,0.4,0.4}
\definecolor{diag1}{rgb}{0.05,0.2,0.6}
\definecolor{diag2}{rgb}{0.55,0.05,0.15}

\begin{figure}[htb]
  \centering
  \begin{subfigure}[b]{0.45\textwidth}
    \begin{tikzpicture}[scale=2.4,baseline]
      \node (nt) at (0,0) {$\color{pnodes}[2,0]_2 [2,0]_2$};
      \node (nJ1) at (-0.5,-0.5) {$\color{pnodes}[2,0]_2 [1,1]_{\tfrac{3}{2}}$};
      \node (nJ2) at (0.5,-0.5) {$\color{pnodes}[1,1]_{\tfrac{3}{2}} [2,0]_2$};
      \node (nj1) at (-1,-1) {${\color{diag1}[2,0]_2} {\color{pnodes}[0,0]_1}$};
      \node (nj2) at (1,-1) {${\color{pnodes}[0,0]_1} {\color{diag2}[2,0]_2}$};
      \node (nh) at (0,-1) {$\color{pnodes}[1,1]_{\tfrac{3}{2}} [1,1]_{\tfrac{3}{2}}$};
      \node (nc1) at (-0.5,-1.5) {${\color{diag1}[1,1]_{\tfrac{3}{2}}} {\color{pnodes}[0,0]_1}$};
      \node (nc2) at (0.5,-1.5) {${\color{pnodes}[0,0]_1}
      {\color{diag2}[1,1]_{\tfrac{3}{2}}}$};
      \node (no) at (0,-2) {${\color{diag1}[0,0]_1} {\color{diag2}[0,0]_1}$};
      \draw[thick,->] (no) edge node[left,anchor=east,xshift=-2mm] {$\aQ$}
      (nc1) (nc1) edge (nj1) (nc2) edge (nh) (nh) edge (nJ1) (nj2) edge (nJ2)
      (nJ2) edge (nt);
      \draw[thick,->] (no) edge node[right,anchor=west,xshift=2mm] {$\bar{\aQ}$}
      (nc2) (nc2) edge (nj2) (nc1) edge (nh) (nh) edge (nJ2) (nj1) edge (nJ1)
      (nJ1) edge (nt);
    \end{tikzpicture}
  \end{subfigure}
  \begin{subfigure}[b]{0.45\textwidth}
    \begin{tikzpicture}[scale=2.4,baseline]
      \node (nt) at (0,0) {$\color{pnodes}[0,0]_3 [0,0]_3$};
      \node (nJ1) at (-0.5,-0.5) {$\color{pnodes}[0,0]_3 [1,1]_{\tfrac{5}{2}}$};
      \node (nJ2) at (0.5,-0.5) {$\color{pnodes}[1,1]_{\tfrac{5}{2}} [0,0]_3$};
      \node (nj1) at (-1,-1) {${\color{diag1}[0,0]_3} {\color{pnodes}[0,2]_2}$};
      \node (nj2) at (1,-1) {${\color{pnodes}[0,2]_2} {\color{diag2}[0,0]_3}$};
      \node (nh) at (0,-1) {$\color{pnodes}[1,1]_{\tfrac{5}{2}} [1,1]_{\tfrac{5}{2}}$};
      \node (nc1) at (-0.5,-1.5) {${\color{diag1}[1,1]_{\tfrac{5}{2}}} {\color{pnodes}[0,2]_2}$};
      \node (nc2) at (0.5,-1.5) {${\color{pnodes}[0,2]_2} {\color{diag2}[1,1]_{\tfrac{5}{2}}}$};
      \node (no) at (0,-2) {${\color{diag1}[0,2]_2} {\color{diag2}[0,2]_2}$};
      \draw[thick,->] (no) edge node[left,anchor=east,xshift=-2mm] {$\aQ$}
      (nc1) (nc1) edge (nj1) (nc2) edge (nh) (nh) edge (nJ1) (nj2) edge (nJ2) (nJ2) edge (nt);
      \draw[thick,->] (no) edge node[right,anchor=west,xshift=2mm] {$\bar{\aQ}$}
      (nc2) (nc2) edge (nj2) (nc1) edge (nh) (nh) edge (nJ2) (nj1) edge (nJ1) (nJ1) edge (nt);
    \end{tikzpicture}
  \end{subfigure}
  \caption{On the left, the ${\color{diag1}A[0,0]} {\color{diag2}A[0,0]}$ multiplet containing $32+32$ degrees
    of freedom. Its superprimary is $\hat{O}^{55}$. On the right, the
    ${\color{diag1}B[0,2]}{\color{diag2}B[0,2]}$ multiplet also containing 32+32 degrees of freedom. Its
    superprimary is $\hat{O}^{ij}$.
  }
  \label{fig:stmult}
\end{figure}
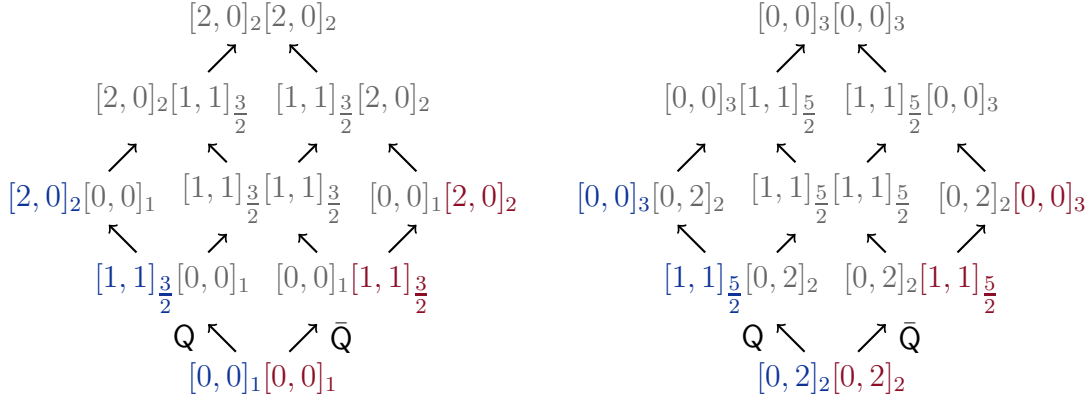

\begin{figure}[tb]
  \centering
  \begin{subfigure}[b]{0.45\textwidth}
    \begin{tikzpicture}[scale=3,baseline]
      \node (nt) at (0.7,1.7) {$\color{pnodes}[1,0]_{\tfrac{3}{2}} [2,1]_{3}$};
      \node (nJ3) at (1.2,1.2) {${\color{pnodes}[0,1]_1} {\color{diag2}[2,1]_{3}}$};
      \node (nJ2) at (0.2,1.2) {$\color{pnodes}[1,0]_{\tfrac{3}{2}} [1,2]_{\tfrac{5}{2}}$};
      \node (nJ1) at (0,1) {$\color{pnodes}[1,0]_{\tfrac{3}{2}} [1,0]_{\tfrac{5}{2}}$};
      \node[align=center] (nh) at (-0.5,0.5) {${\color{diag1}[1,0]_{\tfrac{3}{2}}} {\color{pnodes}[0,1]_{2}}$};
      \node (nj1) at (0.5,0.5) {${\color{pnodes}[0,1]_1} {\color{diag2}[1,0]_{\tfrac{5}{2}}} $};
      \node (nj2) at (0.7,0.7) {${\color{pnodes}[0,1]_1} {\color{diag2}[1,2]_{\tfrac{5}{2}}}$};
      \node (nc) at (0,0) {${\color{diag1}[0,1]_1} {\color{diag2}[0,1]_{2}}$};
      \draw[thick,->] (nc) edge node[left,anchor=west,xshift=2mm] {$\bar\aQ$} (nj1)
      (nj2) edge (nJ3) (nh) edge (nJ1) (nJ2) edge (nt);
      \draw[thick,->] (nc) edge node[right,anchor=east,xshift=-2mm] {$\aQ$}
      (nh) (nj2) edge[transform canvas={xshift=3mm}] (nJ2) (nJ3) edge (nt);
    \end{tikzpicture}
  \end{subfigure}
  \begin{subfigure}[b]{0.45\textwidth}
    \begin{tikzpicture}[scale=3,baseline]
      \node (nt) at (-0.7,1.7) {$\color{pnodes}[2,1]_{3} [1,0]_{\tfrac{3}{2}}$};
      \node (nJ3) at (-1.2,1.2) {${\color{diag1}[2,1]_{3}} {\color{pnodes}[0,1]_1}$};
      \node (nJ2) at (-0.2,1.2) {$\color{pnodes}[1,2]_{\tfrac{5}{2}} [1,0]_{\tfrac{3}{2}}$};
      \node (nJ1) at (0,1) {$\color{pnodes}[1,0]_{\tfrac{5}{2}} [1,0]_{\tfrac{3}{2}}$};
      \node[align=center] (nh) at (0.5,0.5) {${\color{pnodes}[0,1]_{2}} {\color{diag2}[1,0]_{\tfrac{3}{2}}}$};
      \node (nj1) at (-0.5,0.5) {${\color{diag1}[1,0]_{\tfrac{5}{2}}} {\color{pnodes}[0,1]_1}$};
      \node (nj2) at (-0.7,0.7) {${\color{diag1}[1,2]_{\tfrac{5}{2}}} {\color{pnodes}[0,1]_1}$};
      \node (nc) at (0,0) {${\color{diag1}[0,1]_{2}} {\color{diag2}[0,1]_1}$};
      \draw[thick,->] (nc) edge node[left,anchor=east,xshift=-2mm] {$\aQ$} (nj1)
      (nj2) edge (nJ3) (nh) edge (nJ1) (nJ2) edge (nt);
      \draw[thick,->] (nc) edge node[right,anchor=west,xshift=2mm] {$\bar{\aQ}$}
      (nh) (nj2) edge[transform canvas={xshift=-3mm}] (nJ2) (nJ3) edge (nt);
    \end{tikzpicture}
  \end{subfigure}
  \caption{Multiplets ${\color{diag1}B[0,1]}{\color{diag2}A[0,1]}$ and
  ${\color{diag1}A[0,1]}{\color{diag2}B[0,1]}$. They both contain $32+32$ degrees of freedom.}
  \label{fig:stmult2}
\end{figure}


Observe that the conformal dimension of these defect operators is, in some cases, 
lower than that of the corresponding bulk operators, leading to singular terms 
in the dOE. For instance, the dimension of $\hat{O}^{55}$ is 2, whereas the
dimension of $O^{55}$ is 4. A similar behavior occurs in the context of Wilson loops
in 4d $\cN=4$ SYM, where the 1/2-BPS line operator takes the form
\begin{align}
W \sim \tr \cP \exp \im \int \left( A_\tau + \Phi^6 \right) \diff \tau.
\label{eqn:wilsonloop}
\end{align}
In that case, the dOE of the stress tensor superprimary includes a defect
operator of dimension 1, which can be understood as the insertion of $\Phi^6$ in
the line.  Here, we do not have a field realisation of the $\cN = (2,0)$ theory
but $\hat{O}^{55}$ plays an analogous role.

Consider then the dOE for $O^{55}$. From Figures~\ref{fig:stmult}
and~\ref{fig:stmult2} we know some of the defect operators that can appear on
the right-hand side of~\eqref{eqn:doe}. This leads to
\begin{align}
  O^{55}(x) V = \frac{1}{x^4} C_{O \bf{1}}^V V[\mathbf{1}_V]
  + \frac{1}{x^2} C_{O \hat{O}}^V(x, \partial_\sigma) V[\hat O^{55} ]
  + \frac{x_m}{x^2} C_{O\bD}^V(x, \partial_\sigma) V[\bD^m] + \dots
  \label{eqn:o55doe}
\end{align}
The list of defect operators that may appear in this expansion is constrained by
supersymmetry and can be treated systematically, but we do not pursue this
direction further.

Equation~\eqref{eqn:o55doe} can be made more precise.
The coefficients of the defect primaries encode the normalisation of bulk-defect 
correlators as in~\eqref{eqn:doeVev}: 1-point functions such as~\eqref{eqn:StressTensor1Point} 
compute the coefficient of $\mathbf{1}_V$, 2-point functions such 
as~\eqref{eqn:StressTensor2Point} capture the coefficients of other defect 
primaries. Explicitly, $\vev{O^{55}(x) V}$ calculates the defect identity component 
of the dOE, such that
\begin{align}
  C_{O \bf{1}}^V = \frac{h_O}{\pi^3}.
\end{align}
The coefficient of the displacement operator can be found without computing $\vev{O^{55} V[\bD^m]}$ 
explicitly, using the fact that the displacement operator is related to the broken translation symmetry. 
Integrating over the position of $\bD^m$, we can replace it by a derivative:
\begin{align}
  \int \diff^2 \sigma \vev{O^{55}(x) V[\bD^m(\sigma)]} = - \partial^m \vev{O^{55}(x) V}\,.
  \label{eqn:dodoe}
\end{align}
The left hand side is easily computed from~\eqref{eqn:o55doe} and related to $C_\bD$ and 
$C^V_{O \bD}$, while the right hand side is given in terms of $h_O$. Matching coefficients, 
we find
\begin{align}
C^V_{O \bD}(x,\partial_\sigma) = \frac{8 h_O}{\pi^4 C_\bD} \left( 1 + \ldots \right).
\end{align}
By contrast, the coefficient $C_{O \hat{O}}^V$ is not obviously related to the remaining coefficients, 
and thus an independent piece of dCFT data.

\subsection{Constraints from supersymmetry}
\label{sec:dOEsusy}

We conclude this section by sketching an alternative derivation of the results
of Section~\ref{sec:BulkDefectCorrelators}. It turns out
that the dOE provides a simple and elegant way to understand the origin of the linear
relations~\eqref{eqn:1PointSUSYWard} and~\eqref{eqn:swardidentity} without doing
explicit calculations, by reframing them in terms of coefficients of displacement primaries 
in the stress tensor dOE. Indeed, the method we use can in principle be applied far 
more generally to obtain analogous constraints for the remaining dOE coefficients.

To reproduce these results, consider the dOE of $\chi^5$. Following the analysis
of Section~\ref{sec:repsex}, we decompose $\chi^5$ into representations of the preserved
algebra
\begin{align}
[1,1][0,0] \oplus[1,0] [0,1] 
\oplus [0,1] [1,0] \oplus [0,0] [1,1]\,,
\end{align}
which we label $\chi^5_{\alpha_1 \alpha_2}$, $\chi^5_{\alpha_1 \dot\alpha_2}$, 
$\chi^5_{\dot\alpha_1 \alpha_2}$, $\chi^5_{\dot\alpha_1 \dot\alpha_2}$.
We only need the dOE of $\chi^5_{\alpha_1 \dot{\alpha}_2}$, which takes the form
\begin{align}
\label{eqn:chi5doe}
\chi^5_{\alpha_1 \dot\alpha_2} V &= \frac{1}{x^2} C^V_{\chi \bQ} (x, \partial_\sigma)
V[\bQ_{\alpha_1 \dot\alpha_2}] + \ldots
\end{align}
Again, there are other terms that could be included in this expansion, but they
don't play a role in what follows so we ignore them. We also emphasise 
that~\eqref{eqn:chi5doe} is related to the dOE of the stress tensor superprimary 
by supersymmetry.

We can now proceed as in Section~\ref{sec:BulkDefectCorrelators} and find the
constraints imposed by the preserved supersymmetries. Consider first acting with
$\aQ$ on the bulk operator $\chi^5_{\alpha_1 \dot{\alpha}_2}$ to find
\begin{align}
  \aQ \chi = H + j + \partial O\,,
\end{align}
with some coefficients. (The exact expression can be obtained by
restricting~\eqref{eqn:susyTmult} to the relevant representations of the
preserved algebra.) Using the dOE on the right-hand side and focusing on the
defect identity component gives
\begin{align}
\begin{split}
  \big( \aQ \chi(x) \big) V &\sim \big( H(x) + j(x) + \partial O(x) \big) V
  \sim \frac{1}{x^5} \left( C_{H \bf{1}} + C_{j \bf{1}} + C_{\partial O \bf{1}}
  \right) V[\mathbf{1}_V] + \dots 
\end{split}
\label{eqn:QchiBulkChannel}
\end{align}
Note that $C_{j \bf{1}} = 0$ and $C_{\partial O \bf{1}}$ can be obtained
from~\eqref{eqn:dodoe}.  We call this the ``bulk'' channel, since we 
calculate the action of $\aQ$ on $\chi$ before taking the dOE.

The expression~\eqref{eqn:QchiBulkChannel} is to be contrasted with the
``defect'' channel, where we first use~\eqref{eqn:chi5doe} and then apply
$\aQ$.  Clearly, since $\mathbf{1}_V$ is not the variation of anything $\mathbf{1}_V
\neq \aQ(\dots)$, the result does not have an identity component. Consequently, the 
identity component of~\eqref{eqn:QchiBulkChannel} must vanish as well, giving a linear 
constraint equivalent to~\eqref{eqn:1PointSUSYWard} relating the normalisations of the 
stress tensor 1-point functions.

Similarly, \eqref{eqn:swardidentity} can be reproduced by focusing on the scalar displacement 
component of the same equation. The bulk channel gives schematically
\begin{align}
\aQ \chi V &\sim \frac{1}{x^3} \left( C^V_{H \bO} + C^V_{j \bO} + C^V_{\partial O \bO} \right) V[\bO] 
+ \ldots
\end{align}
For the defect channel, we act on~\eqref{eqn:chi5doe} with $\aQ$.
From~\eqref{eqn:dispSusyTrafos}, we see that the variation only leads to
descendants like $\partial \bO$, and no primary. Since equality between defect
and bulk channel must hold at the level of each defect operator, we conclude that
the contribution of the displacement superprimary $\bO$ to the bulk channel must
vanish, and we obtain a linear constraint on the dOE coefficients $C^V_{j \bO},
C^V_{H \bO}, C^V_{\partial O \bO}$, which is equivalent
to~\eqref{eqn:swardidentity}. 
These two relations are only the simplest examples of a much larger set of 
constraints obeyed by the dOE coefficients. Indeed, equating the bulk and defect 
channel of any supercharge acting on any primary dOE at the level of each defect 
operator, it is straightforward to derive further such linear relations. These 
conditions greatly reduce the number of independent coefficients of stress tensor 
dOE coefficients, until we are left with what we could call a super-dOE, i.e. a set 
of dOEs which is fully consistent under the preserved supersymmetry.

\section{Conclusion}
\label{sec:conclusion}

In this paper, we initiate the application of defect CFT techniques to describe 
surface operators of the 6d $\cN = (2,0)$ theory, that is, we apply the ideas and
tools of CFT to study local operator insertions into the 1/2-BPS plane.
An important insertion is the displacement operator~\eqref{eqn:dispop} which
literally deforms the plane, but there are also other defect operators
corresponding to inserting bulk operators near the defect---they are captured by
the dOE~\eqref{eqn:doe}.

One of our results is the classification of unitary multiplets of $\osp(4^*|2)
\oplus \osp(4^*|2)$, the algebra preserved by a 1/2-BPS defect, in 
Section~\ref{sec:unitarity}. These multiplets were not classified before to the best of our
knowledge\footnote{%
  Note added: the same classification of unitary multiplets of $\osp(4^*|2)$ was
  also done in~\cite{Agmon:2020pde}, which appeared as this paper was finalised.
}
and are the building blocks for discussing other aspects of the dCFT,
like its spectrum, the OPE of defect operators and the dOE.
In this work we focus on the dOE, but it would also be interesting to pursue
these other directions, for instance using the tools of conformal
bootstrap~\cite{Liendo:2012hy}.

There are two important applications of the dOE~\eqref{eqn:doe} in our analysis: in
Section~\ref{sec:repsex} we use it to find new defect operators and in
Section~\ref{sec:dOEsusy} we sketch how it makes the preserved symmetries
manifest.

First, we use it to give the example of how the bulk stress tensor multiplet
decomposes into defect multiplets. There are of course the operators $\bD$, $\bQ$
and $\bO$ of the displacement multiplet, but also other defect multiplets whose
operator content is shown in Figure~\ref{fig:stmult} and~\ref{fig:stmult2}.
Although we focus on the stress tensor multiplet, this analysis could also be
applied to any other multiplet of the $\cN = (2,0)$ theory. In addition to the
multiplets presented above, the dOE can include additional terms, and it
would be interesting to obtain the selection rules as was done for 4d $\cN = 4$
SYM~\cite{Liendo:2016ymz}, by treating systematically all the superconformal
Ward identities.

The important aspect of this decomposition of bulk operators is that it is
convergent. In particular, it encodes all the information of the bulk OPE, which
opens the possiblity of studying the $\cN=(2,0)$ theory from the point of view
of a 2d defect CFT. This direction could lead to additional constraints on the
bulk theory, since the defect operators are not a trivial rewriting of those in the
bulk. This is manifested for instance by the appearance of divergences in the
dOE of $O^{55}$~\eqref{eqn:o55doe}.

Instead, the dOE captures some important reorganisation of degrees of freedom in
the dCFT.  For instance, in the expansion of the bulk operator $O^{55}$~\eqref{eqn:o55doe}
we find a defect operator which is of dimension 2 and therefore marginal (we
expect it to be marginally irrelevant). The analogous expansion of the
superprimary of the stress tensor multiplet is well understood in the context of
Wilson loops in 4d $\cN = 4$ SYM: using the definition of the 1/2-BPS Wilson
loop~\eqref{eqn:wilsonloop} the marginal operator there corresponds to inserting
$\Phi^6$ into the line defect~\cite{Polchinski:2011im}. Here the interpretation
is similar: inserting the analog of $\hat{O}^{55}$ in the non-supersymmetric
surface operator triggers an RG flow which comes to a stop when $\hat{O}^{55}$
becomes marginal at the conformal fixed point, which is the 1/2-BPS surface
operator. This flow is verified in holography~\cite{Drukker:2020swu} and should
hold more generally for all $\cN = (2,0)$ theories.

A second use of the dOE is to make the preserved symmetries manifest. As we
sketch in Section~\ref{sec:dOEsusy}, we can explain the origin of the
relation between $h_T$ and $C_\bD$~\eqref{eqn:result} simply by looking at the structure of multiplets of
defect operators. This is to be contrasted with the derivation of
Section~\ref{sec:BulkDefectCorrelators}, where the relation is the result of a
calculation and not obvious from the outset. We believe this approach could
shed light on determining the minimal amount of supersymmetry required to
prove~\eqref{eqn:result}, that is whether it also holds
for defects of the $\cN = (1,0)$ theory, and more generally what are the
necessary conditions to prove the conjecture of~\cite{Lewkowycz:2013laa}.

In addition to the methods, the result~\eqref{eqn:result}
and the corresponding relation between the anomaly
coefficients~\eqref{eqn:anomalyresult} are themselves interesting. In the
context of Wilson loops, $C_\bD$ was shown to appear in the Bremsstrahlung
function~\cite{correa:2012at} and $h_T$ both in the radiation emitted by a quark
undergoing constant acceleration and the entanglement entropy associated with
Wilson lines~\cite{Lewkowycz:2013laa}. While these calculations can be
generalised to the case of surface operators, they do not give a finite answer:
as shown in Section~\ref{sec:anomaly} respectively inserting displacement
operators in the defect or introducing a stress tensor in its vicinity leads to
a logarithmic divergence caused by an anomaly. To obtain a finite result, one
can define a renormalised surface operator
\begin{align}
  V_{\epsilon} = \epsilon^{-\int \cA_\Sigma \diff^2 \sigma} V\,,
\end{align}
so that $V_{\epsilon}$ effectively acquires a dimension ($\cA_\Sigma$ is defined
in~\eqref{eqn:anomaly}). The interpretation of $C_\bD$ and $h_T$ are then as the
anomaly coefficients $b_1$ and $b_2$, which are the relations~\eqref{eqn:b1cd},
\eqref{eqn:CTb2relation} (also \eqref{eqn:cco} between $c$ and $C_\bO$). The net result
of the relations between the anomaly coefficients~\eqref{eqn:anomalyresult} is
that the anomaly depends on the geometry only through the combination $(H^2 + 4
\tr P) - (\partial n)^2$ (defined in~\eqref{eqn:GaussCodazzi}), while the anomaly term
$\RicciScalar^\Sigma$ integrates to a topological invariant, the Euler
characteristic of the surface $\Sigma$.  We note that for some classes of BPS operators,
$H^2$ and $(\partial n)^2$ are related and either cancel out or give interesting
quantities. A simple example is the uplift of the BPS Wilson loops
of~\cite{zarembo:2002an} for which the anomaly vanishes, but more examples will
be presented elsewhere~\cite{S3-surfaces}.

Finally, there are other interesting directions which we haven't explored in
this paper. For the Wilson line, a point of confluence between different
techniques is the cusp, whose anomalous dimension at small angles is related to
the Bremsstrahlung function~\cite{correa:2012at} and can be calculated using
integrability~\cite{Drukker:2012de,Correa:2012hh,Gromov:2015dfa} and
supersymmetric localization~\cite{Pestun:2009nn}.  Its analog here are conical
singularities which exhibit a peculiar $\log^2{\epsilon}$ divergence,
see~\cite{klebanov:2012yf,myers:2012vs,bueno:2015lza,dorn:2016bkd,Drukker:2020dcz}.
The coefficient of the divergence is entirely fixed by the behavior of the
surface near the singularity, so it is natural to consider an operator inserting
a conical singularity and to try and find its interpretation in the dCFT.

Another possibility is to study further the OPE for BPS operators.
The $\cN = (2,0)$ theory contains a sector isomorphic to a chiral
algebra~\cite{Beem:2014kka} which can be used to calculate for instance the
3-point functions of 1/4-BPS local operators. For 4d $\cN = 2$ SCFTs, it was
shown in~\cite{Cordova:2017mhb} that the supercharges defining the cohomology
are compatible with $\cN = (2,2)$ surface defects, and it would be interesting
to extend their construction to the $\cN = (2,0)$ theory with 1/2-BPS surface
defects. This could lead to exact results for a sector of the dOE and defect
OPE.

It would also be interesting to study BPS operators in the context of the AGT
correspondence. At large $N$ one can use holography to calculate the expectation
values, in the presence of the defect, of operators in the traceless symmetric
representation of $\sof(5)_R$~\cite{Corrado:1999pi}, which contains in
particular $O^{IJ}$ in the stress tensor multiplet. Since the the AGT
correpondence can be used to calculate the expectation value of the stress
tensor~\cite{Chalabi:2020iie}, it might also calculate expectation values for
this larger class of operators at finite $N$.

\subsection*{Acknowledgements}

We would like to thank
Chris Herzog and
Marco Meineri
for interesting discussions. The work of ND is supported by the STFC grants
ST/P000258/1 and ST/T000759/1.
MT acknowledges the support of the Natural Sciences and Engineering Research Council 
of Canada (NSERC).  
Cette recherche a \'et\'e financ\'ee par le Conseil de recherches en sciences naturelles 
et en g\'enie du Canada (CRSNG).

\appendix

\section{Conventions}
\label{sec:convention}

We make use of the following indices:
$$
\label{indices}
  \begin{tabular}{l | l}
    Index & Usage \\ \hline
    $\mu = 1, \dots, 6$ & 6d spacetime coordinates\\
    $m = 1,\dots,4$ & coordinates transverse to the surface $x^m$\\
    $a = 1,2$ & worldsheet coordinates $\sigma^a$\\
    $\alpha\ (\dot{\alpha}) = 1, \dots, 4$ & 6d chiral (antichiral) spinors\\
    $I = 1, \dots, 5$ & R-symmetry vectors\\
    $i = 1, \dots, 4$ & preserved R-symmetry\\
    $\check{\alpha} = 1,\dots,4$ & R-symmetry spinors\\
  \end{tabular}
$$
We work in Minkowski space with mostly positive signature.
Below we detail the properties of gamma matrices arising in the
$\mathfrak{osp}(8^*|4)$ algebra. More details can be found
in~\cite{claus:1997cq} and references therein.

\subsection{Spinors and \texorpdfstring{$\gamma$}{gamma}-matrices}

We suppress contracted spinor indices in the main text. We use the NW-SE spinor index
convention, so that
\begin{align}
  \varepsilon \psi \equiv \varepsilon^{\alpha \check{\alpha}} \psi_{\alpha
    \check{\alpha}},
\end{align}
where $\alpha$ is the index of a chiral 6d spinor ($\dot{\alpha}$ for
antichiral) and $\check{\alpha}$ that of an R-symmetry spinor. These indices are
raised and lowered by the charge conjugation matrices $c, \Omega$, which obey
\beq
  c^\dagger c = c^{\alpha \dot{\alpha}} c_{\dot{\alpha} \beta}
  = \delta^{\beta}_{\alpha}\,, 
  \qquad
  c^* c^T = c^{\dot{\alpha} \alpha} c_{\alpha \dot{\beta}}
  = \delta^{\dot{\beta}}_{\dot{\alpha}}\,, 
  \qquad
  \Omega^\dagger \Omega =
  \Omega^{\check{\alpha} \check{\beta}} \Omega_{\check{\beta} \check{\gamma}}
  = \delta^{\check{\alpha}}_{\check{\gamma}}\,.
\eeq
We also make use of two types of $\gamma$-matrices:
6d chiral $\tensor{\left( \gamma_\mu \right)}{_\alpha^{\dot{\beta}}}$
(antichiral $\tensor{\left( \bar{\gamma}_\mu \right)}{_{\dot{\alpha}}^\beta}$)
and 5d $\tensor{(\check{\gamma}_I)}{_{\check{\alpha}}^{\check{\beta}}}$
associated to R-symmetry. Their algebra is
\beq
  \bar{\gamma}_\mu \gamma_\nu + \bar{\gamma}_\nu \gamma_\mu = 2 \eta_{\mu\nu}\,, 
  \qquad
   \gamma_\mu \bar{\gamma}_\nu + \gamma_\nu \bar{\gamma}_\mu = 2 \eta_{\mu\nu}\,, 
   \qquad
    \left\{ \check{\gamma}_I, \check{\gamma}_J \right\} = 2 \delta_{I J}\,.
\eeq
The chiral and antichiral representations are related through
\beq
  \bar{\gamma}_\mu^\dagger = \gamma_0 \bar{\gamma}_\mu \gamma_0
  = \gamma^\mu\,,
\eeq
and the matrices are antisymmetric
\beq
  (\gamma_\mu c) = - \left(\gamma_\mu c \right)^T\,, \qquad
  \left( \bar{\gamma}_\mu c^T \right)
  = - \left( \bar{\gamma}_\mu c^T \right)^T\,, \qquad
  (\check{\gamma}_I \Omega) = - \left(\check{\gamma}_I \Omega \right)^T\,.
\eeq
Because the matrices are chiral, they also satisfy
\beq
  \gamma_{012345}
  = I\,, 
  \qquad
  \bar{\gamma}_{012345}
  = -I\,,
  \qquad
  \check{\gamma}_{12345} = I\,,
\eeq
with $\gamma_{\mu\nu \dots \rho} \equiv \gamma_{[\mu} \bar{\gamma}_{\nu} \dots
\gamma_{\rho]}$ the antisymmetrised product of $\gamma$-matrices.

A representation of this algebra is given by
\begin{align}
  \gamma_0 &= \bar{\gamma}_0 = i I_2 \otimes I_2\,,&\qquad
  \gamma_1 &= -\bar{\gamma}_1 = -i \sigma_1 \otimes I_2\,,&\qquad
  \gamma_2 &= -\bar{\gamma}_2 = -i \sigma_2 \otimes I_2\,,
  \nonumber\\
  \gamma_3 &= -\bar{\gamma}_3 = i \sigma_3 \otimes \sigma_1\,,&\qquad
  \gamma_4 &= -\bar{\gamma}_4 = i \sigma_3 \otimes \sigma_2\,,&\quad
  \gamma_5 &= -\bar{\gamma}_5 = -i \sigma_3 \otimes \sigma_3\,,
  \nonumber\\[0.2\baselineskip]
  \check{\gamma}_1 &= \sigma_1 \otimes \sigma_2\,,\quad
  \check{\gamma}_2 = \sigma_2 \otimes \sigma_2\,,\hskip-6mm&
  \check{\gamma}_3 &= \sigma_3 \otimes \sigma_2\,,\quad
  \check{\gamma}_4 = I_2 \otimes \sigma_1\,,\hskip-6mm&
  \check{\gamma}_5 &= I_2 \otimes \sigma_3\,,
  \nonumber\\[0.2\baselineskip]
  c &= -c^T = \sigma_1 \otimes i \sigma_2\,,&\qquad
  \Omega &= i \sigma_2 \otimes I_2\,.\hskip-3cm
  \label{eqn:gammarep}
\end{align}

\section{Weyl anomaly for surface operators}
\label{app:anomaly}

Surface operators in CFT typically suffer from UV divergences which make their
expectation value ill-defined. Up to power-law divergences (which can be removed
by appropriate counterterms) their expectation value takes the form
\beq
  \log{\vev{V_\Sigma}} \sim
  \log{\epsilon} \int_\Sigma \vol_\Sigma \mathcal{A}_\Sigma  +
  \fin,
\eeq
where $\epsilon$ is a regulator, $\Sigma$ is the surface (in this paper we take
the plane) and $\mathcal{A}_\Sigma$ is known as the anomaly density. 

This conformal anomaly is constrained by the Wess-Zumino condition to take the form
\beq
\mathcal{A}_\Sigma = \frac{1}{4 \pi} \left[
  a \RicciScalar^{\Sigma} + b_1 \tr{\tilde{\IIFundForm}^2} + b_2 \tr{W} +c(\partial n)^2
  \right].
\eeq
The conformal invariants of this expression are
\begin{itemize}
\item[]
\makebox[2.3cm][r]{$\RicciScalar^{\Sigma}$:} The Ricci scalar on $\Sigma$.
\item[]
  \makebox[2.3cm][r]{$\tr{\tilde{\IIFundForm}^2}$:} The square of the traceless
part of the second fundamental form.
\item[]
\makebox[2.3cm][r]{$\tr{W}$:} $W$ is the pullback of the Weyl tensor.
\item[]
\makebox[2.3cm][r]{$(\partial n)^2$:} The norm of the variation of the coupling
to R-symmetry.
\end{itemize}
The exact definition of these invariants in our conventions can be found in Appendix B
of~\cite{Drukker:2020dcz}. Here we use the definition of the second
fundamental form in~\eqref{eqn:DispIIFundForm}
\beq
  \IIFundForm_{a b}^{\mu} = \left(\partial_a \partial_b x^\lambda 
  + \partial_a x^\rho \partial_b x^\sigma \Gamma^\lambda_{\rho\sigma}\right) 
  \left(\delta_\lambda^\mu - g_{\kappa\lambda}\partial^c x^\kappa \partial_c
  x^\mu \right),
  \label{eqn:IIFundForm}
\eeq
which for $\Gamma^\lambda_{\rho \sigma} = 0$ and $x^m = \xi^m$ reduces
to~\eqref{eqn:DispIIFundForm}. It can be decomposed into its trace, $H^\mu$, and
its traceless part, $\tilde{\IIFundForm}^\mu_{ab}$, which are related by the
Gauss-Codazzi equation to other invariants
\beq
  \left( H^2 + 4 \tr{P} \right) =
  2 \RicciScalar^{\Sigma} + 2 \tr{\tilde{\IIFundForm}^2} - 2 \tr{W}\,.
  \label{eqn:GaussCodazzi}
\eeq

\section{Conformal Ward identities for defect correlators}
\label{app:ConformalWardIdentities}

In this appendix, we derive explicit expressions for the structure of the expectation values of stress tensor primaries 
in the presence of a flat conformal surface defect. Up to overall normalisation constants, which we 
further constrain in Section~\ref{sec:BulkDefectCorrelators} using supersymmetry, these correlators 
are completely fixed by the bosonic symmetries (conformal and R-symmetry) preserved by the defect. 
We consider both defects with an insertion of a single primary of the displacement operator multiplet, 
and defects without such insertions. For brevity, we do not give an exhaustive list of such correlators 
and instead focus on those we require in the main text. More specifically, we
compute only the expectation values of the primaries in the stress tensor
multiplet, and some 2-point functions involving low-level 
primaries, namely $O^{IJ}, \chi^I_{\alpha \check\alpha}, H_{\lambda\mu\nu}^I$ in the stress tensor, 
and $\bO^i, \bQ_{\alpha\check\alpha}$ in the displacement multiplet. The remaining correlators can of 
course be calculated using the same method. 

We proceed in two steps. First, 
we fix the dependence on $\sigma$ and $x$ by implementing the Ward identities associated with the 
conformal symmetry preserved by the defect as well as transverse rotational symmetry. For clarity, in this
calculation we suppress the R-symmetry indices of the operators and leave the scaling dimensions general.
Indeed, as much of the kinematics is easily generalised to defects of dimension $p$ in arbitrary spacetime 
dimension $d=p+q$, we state the more general result wherever we can do so without obscuring the 
results we presently need. Secondly, we fix the R-symmetry tensor structure of these correlators by demanding 
invariance under the residual $\sof(4)_R$ symmetry. Throughout, we denote generic operators in the bulk
$\cO$ and on the defect $\hat\cO$.

Many of the kinematical results have been obtained by different methods in the past. In particular, the 
embedding space formalism allows for the efficient computation of bosonic correlators~\cite{Billo:2016cpy}.
However, it is not straightforwardly applicable to correlators involving fermions. 

\subsection{Defect without insertions}
\label{subsec:ConformalNoInsertions}

We want to solve the constraints that the residual conformal symmetry places on
expectation values of the form $\vev{\cO V}$ with $\cO$ a bulk
operator of scaling dimension $\Delta$. The representation of the conformal
algebra~\eqref{eqn:ospconformalalgebra} acting on $\cO$ is given in terms of
the representation of $\cO$ under Lorentz transformations $S_{\mu \nu}$ and is
\begin{equation}
  \begin{gathered}
    \label{eqn:repconf}
    \aP_\mu = \partial_\mu, \qquad
    \aM_{\mu\nu} = 2 x_{[\mu} \partial_{\nu]} + S_{\mu\nu}, \qquad
    \aD = -x^\mu \partial_\mu - \Delta\,,\\
    \aK_\mu = x^2 \partial_\mu - 2 x_\mu (x^\nu \partial_\nu
    + \Delta) + 2 x^\nu S_{\nu \mu}.
  \end{gathered}
\end{equation}
Treating separately the coordinates along the plane $\sigma^a$ and tranverse
$x^m$, translation invariance on the plane implies that $\vev{\cO(\sigma, x) V}$
is a function of $x^m$ only. The other Ward identities can be cast into the form:
\begin{align}
\label{eqn:ConformalWardIdentities1Point}
\begin{split}
0 &= S_{ab} \vev{\cO V }\,, \\
0 &= \left( x^m \partial_m + \Delta \right) \vev{\cO V }\,, \\
0 &= x^m S_{am} \vev{\cO V}\,, \\
0 &= \left( x_m \partial_n - x_n \partial_m \right) \vev{\cO V} + S_{mn}
\vev{\cO V}\,.
\end{split}
\end{align}

These constraints are now straightforwardly solved. We focus on scalars $O$, vectors $j_\mu$, selfdual 
3-forms $H_{\lambda\mu\nu}$ and traceless symmetric 2-tensors $T_{\mu\nu}$, as
operators of those types 
make up the bosonic degrees of freedom of the stress tensor multiplet, while the correlators of fermionic 
operators with a scalar defect vanish identically.

For a Lorentz scalar $O$, all $S_{\mu\nu}$ vanish and the conformal Ward 
identities~\eqref{eqn:ConformalWardIdentities1Point} are immediately solved to give 
\begin{align}
\label{eqn:1PointScalar}
\vev{ O(\sigma, x) V } = \frac{h_O}{x^{\Delta}}\,,
\end{align}
with $h_O$ an as yet undetermined constant. 

The transformation law for a vector reads 
\begin{align}
\label{eqn:VectorLorentzTransformation}
  \left( S_{\mu\nu} j \right)_\rho
  = \delta_{\mu\rho} j_\nu - \delta_{\nu\rho} j_\mu\,,
\end{align}
which, plugged into~\eqref{eqn:ConformalWardIdentities1Point} eventually leads 
to\footnote{More generally, for a $p$-dimensional defect in a spacetime of dimension $d = p+q$, one obtains 
\begin{align*}
\vev{j_a(x) V} = 0\,, \qquad (q-2) \vev{j_m(x) V} = 0\,.
\end{align*}
Indeed, for $q=2$, the transverse components of $j$ can take the form
\begin{align*}
\vev{j_m(x) V } \sim \frac{\epsilon_{mn} x^n}{{x}^{\Delta+1}}\,,
\end{align*}
which is compatible with conservation.}
\begin{align}
\label{eqn:1PointVector}
\vev{j_a V} = \vev{j_m V} = 0.
\end{align}

For higher spin bosonic operators, each Lorentz index separately transforms 
as~\eqref{eqn:VectorLorentzTransformation}. For a 3-form $H_{\lambda\mu\nu}$, 
the Ward identities \eqref{eqn:ConformalWardIdentities1Point} imply that the only 
components with nonvanishing expectation value 
in the presence of $V$ are $H_{abm}$ and $H_{lmn}$, and furthermore restricts the 
available terms for their one-point functions to
\begin{align}
\label{eqn:3Form1Point}
\vev{H_{abm}(x) V} \sim \frac{\epsilon_{ab}x_m}{x^{\Delta+1}}\,, \qquad
\vev{H_{lmn}(x) V} \sim \frac{\epsilon_{lmnp}x^p}{x^{\Delta+1}}\,.
\end{align}
In this work, we are concerned with 3-forms which come with a selfduality condition, which 
serves to relate the proportionality constants in~\eqref{eqn:3Form1Point}. We are left with
\begin{align}
\label{eqn:3FormSelfDual1Point}
\vev{H_{abm}(x) V} = h_H \frac{\epsilon_{ab}x_m}{x^{\Delta+1}}\,, \qquad
\vev{H_{lmn}(x) V} = h_H \frac{\epsilon_{lmnp}x^p}{x^{\Delta+1}}\,.
\end{align}
Lastly, we repeat the same analysis for a symmetric traceless 2-tensor. Exactly the same line 
of argument as above yields 
\begin{align}
\label{eqn:1PointSym2Tensor}
\begin{split}
\vev{T_{ab}(x) V} &= \frac{h_T}{{x}^{\Delta}} \delta_{ab}\,, \qquad
\vev{T_{am}(x) V}=0\,, \\
\vev{T_{mn}(x) V} &= \frac{h_T}{{x}^{\Delta+2}} \left( 2 x_m x_n - x^2
\delta_{mn}\right)\,.
\end{split}
\end{align}

We are now in a position to construct the correlator of $V$ with any bosonic primary in the stress 
tensor multiplet. To that end, recall that, under the unbroken $\sof(5)_R$, $O^{IJ}$ and $H^I_{\lambda\mu\nu}$ 
transform as a symmetric traceless 2-tensor and a vector, respectively, while the stress tensor 
$T_{\mu\nu}$ is an R-symmetry singlet.\footnote{The R-symmetry current $j^{IJ}_\mu$ transforms 
as an antisymmetric tensor, but as seen above, its 1-point function vanishes identically regardless of 
the R-symmetry structure.} 
Without explicitly applying the Ward identities associated 
with the preserved $\sof(4)_R$, we can fix the R-symmetry structure of the 1-point functions by writing 
down the available terms and, for $O^{IJ}$, implementing tracelessness. Plugging in the correct scaling 
dimensions $\Delta_O=4$, $\Delta_H = 5$, and $\Delta_T  = 6$, we find the only nonvanishing 1-point 
functions of stress tensor primaries in the presence of $V$
are~\eqref{eqn:StressTensor1Point}.

\subsection{Defect with an insertion}

We now repeat the above discussion for correlators $\vev{\cO(\sigma, x)
V[\hat\cO(\sigma')] }$
involving a defect with an insertion of a displacement multiplet primary. The kinematical analysis is more 
involved than, but technically very similar to, the previous subsection. We use translation invariance to 
center $\hat{\cO}$ at $\sigma'=0$ and suppress the arguments of $\cO(\sigma, x)$. The conformal 
Ward identities may be cast into the form:
\begin{align}
\label{eqn:ConformalWardIdentities2Point}
\begin{split}
0 &= \left( \left( \sigma_a \partial_b - \sigma_b \partial_a \right) +
\hat{S}_{ab} + S_{ab} \right) \vev{\cO V[\hat\cO] }\,, \\
0 &= \left( \left( x_m \partial_n - x_n \partial_m \right) + \hat{S}_{mn} +
S_{mn} \right) \vev{\cO V[\hat\cO] }\,, \\
0 &= \left( \sigma^a \partial_a + x^m \partial_m + \Delta + \hat\Delta \right)
\vev{\cO V[\hat\cO] }\,, \\
0 &= \left( 2x^m S_{am} + 2 \sigma^b S_{ab} + 2\hat{\Delta} \sigma_a + (\sigma^2
+ x^2) \partial_a \right) \vev{\cO V[\hat\cO] }\,.
\end{split}
\end{align}
For the simplest case of a scalar $\bO$ on the defect and a scalar $O$ in the bulk, \eqref{eqn:ConformalWardIdentities2Point} become particularly simple, and 
imply\footnote{In particular, inserting for $\bO$ the defect identity operator $\mathbf{1}_V$, 
we recover the form of~\eqref{eqn:1PointScalar}, as expected.}
\begin{align}
\vev{O(\sigma, x) V[\bO] } = \frac{C_{O \bO}}{{x}^{\Delta - \hat\Delta} (\sigma^2 + x^2)^{\hat\Delta}},
\end{align}
with $C_{\bO O}$ some normalisation constant.

For a defect scalar $\bO$ and a bulk vector $j_\mu$ we obtain:
\begin{equation}
  \label{eqn:2PointVectorScalar}
  \begin{aligned}
  \vev{j_a (\sigma, x) V[\bO] } &=
  \frac{C_{j \bO} \sigma_a}{{x}^{\Delta - \hat\Delta - 1} (\sigma^2 + x^2)^{\hat\Delta + 1} }\,, \\
  \vev{j_m (\sigma, x) V[\bO] } &=
  \frac{C_{j \bO} (x^2 - \sigma^2) x_m}{ 2 {x}^{\Delta - \hat\Delta + 1}
  (\sigma^2 + x^2)^{\hat\Delta + 1}}\,.
  \end{aligned}
\end{equation}
Indeed, these correlators are exactly the same for defects of generic dimension and codimension. It is 
easily checked that~\eqref{eqn:2PointVectorScalar} is compatible with 
conservation of $j$ in the bulk if and only if $\Delta = d-1$ and $\hat\Delta = p$, which is indeed satisfied by 
the displacement superprimary $\bO^i$ and the bulk R-symmetry current $j^{IJ}_\mu$. 
The conservation equation
\begin{align}
\label{eqn:ConservationEquation}
\partial_\mu \vev{j^\mu(\sigma, x) V[\bO]}= \vev{V[\bO(\sigma) \bO(0)]}, 
\end{align}
then allows us to fix $C_\bO$ in terms of $C_{\bO j}$ in equation~\eqref{eqn:Conservation}. 
For the remaining required bosonic correlator, consider a defect scalar $\bO$ and a bulk 
3-form $H_{\lambda\mu\nu}$. The conformal Ward identities \eqref{eqn:ConformalWardIdentities2Point} 
imply that the only components of the correlator that do not vanish identically are
\begin{equation}
  \begin{aligned}
    \vev{H_{abm}(\sigma, x) V[\bO] } &=
    \frac{h_H \epsilon_{ab} x_m}{x^{\Delta-\hat\Delta+1} (\sigma^2 + x^2)^{\hat\Delta}}\,,  \\
    \vev{H_{lmn}(\sigma, x) V[\bO] } &=
    \frac{h_H \epsilon_{lmnp}x^p}{x^{\Delta-\hat\Delta+1}
    (\sigma^2+x^2)^{\hat\Delta}}\,,
  \end{aligned}
\end{equation}
where, as for the 1-point function, we have used the selfduality of $H_{\lambda\mu\nu}$ to relate the 
two normalisation constants.
Lastly, we compute the only correlator of fermions that we require in this paper. Consider a bulk 
chiral spinor $\chi_\alpha$ and a defect chiral spinor $\bQ_{\alpha}$.%
\footnote{
  Since ultimately we are interested in a defect operator defined in terms of a
  chiral fermionic bulk current, we take $\bQ$ to transform as a spinor under both
  parallel and transverse rotations, and consider only chiral objects.
} 
Their transformation laws are familiar:
\begin{equation}
  \begin{aligned}
    \left( S_{\mu\nu} \chi \right)_\alpha =
    \frac{1}{2} \left( \gamma_{\mu\nu}
    \right)\indices{_\alpha^\beta}\chi_\beta\,, \qquad
    \left( S_{ab} \bQ \right)_\alpha =
    \frac{1}{2} \left( \gamma_{ab} \right)\indices{_\alpha^\beta}\bQ_\beta\,,
    \qquad
    \left( S_{mn} \bQ \right)_\alpha =
    \frac{1}{2} \left( \gamma_{mn} \right)\indices{_\alpha^\beta}\bQ_\beta\,.
  \end{aligned}
\end{equation}
In order to apply the Ward identities~\eqref{eqn:ConformalWardIdentities2Point}, we expand 
$\vev{\chi_\alpha V[\bQ_\beta]}$ in terms of antisymmetrised products of gamma 
matrices.
The only such matrices with the appropriate chirality properties are $\gamma^\mu$ and 
$\gamma^{\mu\nu\rho}$ (we can omit $\gamma^{\mu\nu\rho\sigma\tau}$ since it is related to 
$\gamma^\mu$ by duality):
\begin{align}
\vev{\chi_\alpha V[\bQ_\beta]} = a_\mu \left(\gamma^\mu c\right)_{\alpha\beta} + \frac{1}{3!} 
b_{\lambda\mu\nu} \left(\gamma^{\lambda\mu\nu} c\right)_{\alpha\beta}.
\end{align}
Writing out and simplifying the conformal Ward identities explicitly then leads to 
\begin{align}
\vev{\chi_\alpha(\sigma,x) V[\bQ_\beta]} = \frac{c_{\chi\bQ} \left[(\sigma_a \gamma^a + x_m \gamma^m)c\right]_{\alpha\beta}}{x^{\Delta - \hat\Delta} \sqrt{\sigma^2 + x^2}^{1+2\hat\Delta}}.
\end{align}

Having completed the kinematic analysis, we can now restore the R-symmetry structure in order to 
construct the full bulk-defect 2-point functions. The Ward identities associated with the generators 
of $\sof(4)_R$ decouple from the kinematics, and therefore take a purely
algebraic form (with $R, \hat{R}$ the representations of $\cO, \hat{\cO}$)
\begin{align}
\label{eqn:RSymmetryWardIdentity}
0 = \left( R^{ij} + {\hat R}^{ij} \right) \vev{\cO V[\hat\cO]}.
\end{align}
Among the bosonic 2-point functions we consider, the only nonvanishing ones are
(we again suppress coordinate dependence and Lorentz indices):
\begin{align}
\vev{O^{i5}V[\bO^j]} \sim \delta^{ij}\,, \qquad
\vev{j^{i5}V[\bO^j]} \sim \delta^{ij}\,, \qquad
\vev{H^{i}V[\bO^j]} \sim \delta^{ij}\,.
\end{align}
To restore the correct R-symmetry structure of the fermionic 2-point function, recall that 
$\chi^I_{\alpha \check\alpha}$ transforms in the tensor product of the vector and spinor representation 
of $\sof(5)_R$ and is subject to a constraint $\check\gamma_I \chi^I = 0$, while $\bQ_{\alpha \check\alpha}$ 
transforms as an ordinary R-symmetry spinor but obeys a constraint $\Pi_+ \bQ = 0$ mixing Lorentz and 
R-symmetry. Since we only need the correlator involving $\chi^5_{\check\alpha}$, we make the ansatz
\begin{align}
\vev{\chi^5_{\check\alpha} \bQ_{\check\beta}} \sim \left(\check\gamma^5\right)_{\check\alpha \check\beta},
\end{align}
which is indeed compatible with~\eqref{eqn:RSymmetryWardIdentity}.

With the kinematical data and R-symmetry structure in hand, we can now assemble the full 2-point functions. 
Plugging in the correct defect operator scaling dimensions $\Delta_\bO = 2$ and
$\Delta_\bQ = 5/2$, we obtain~\eqref{eqn:StressTensor2Point}.

\section{Algebras}
\label{sec:ospalgebra}

In this appendix we collect some results on the algebras $\mathfrak{osp}(8^*|4)$
and $\osp(4^*|2) \oplus \osp(4^*|2)$. For a general reference on Lie superalgebra,
see~\cite{kac:1977qb, frappat:1996pb} and references therein.

\subsection{The algebra \texorpdfstring{$\mathfrak{osp}(8^*|4)$}{osp^*(8|4)}}

The quaternionic orthosymplectic algebra $\mathfrak{osp}(8^*|4) = D(4,2)$ is a
6d superconformal algebra containing 38 bosonic and 32 fermionic
generators.\footnote{%
  More precisely, it is a real form of $D(4,2)$ given by $\aP_\mu^\dagger = \aK^\mu$
  (which also implies $(\aQ_{\alpha \check{\alpha}})^\dagger = \aS^{\alpha
    \check{\alpha}}$) and compatible with radial quantisation in Euclidean
    space. Hermitean generators can be obtained by redefining all generators
    $\aP \to i \aP$.
}
Its bosonic part $\sof(2,6) \oplus \sof(5)$ contains a 6d conformal algebra
\begin{equation}
  \begin{aligned}
  \left[ \aM_{\mu\nu}, \aM_{\rho \sigma} \right]
  &= 2 \eta_{\sigma [\mu} \aM_{\nu] \rho}
  - 2 \eta_{\rho [\mu} \aM_{\nu] \sigma}\,,\qquad&
  \left[ \aP_\mu, \aK_\nu \right]
  &= 2 \left( \aM_{\mu\nu} + \eta_{\mu\nu} \aD \right)\,,\\
  \left[ \aM_{\mu \nu}, \aP_{\rho} \right]
  &= 2 \aP_{[\mu} \eta_{\nu] \rho}\,,& 
  \left[ \aM_{\mu \nu}, \aK_\rho \right]
  &= 2 \aK_{[\mu} \eta_{\nu] \rho}\,,\\
  \left[ \aD, \aP_\mu \right] &= \aP_\mu\,,&
  \left[ \aD, \aK_\mu \right] &= -\aK_\mu\,,
  \end{aligned}
  \label{eqn:ospconformalalgebra}
\end{equation}
along with an $\sof(5)$ R-symmetry
\begin{align}
  \begin{gathered}
  \left[ \aR_{IJ}, \aR_{KL} \right]
  = 2 \delta_{K [I} \aR_{J] L} - 2 \delta_{L [I}
  \aR_{J] K}\,.
  \end{gathered}
\end{align}
The fermionic generators $\aQ$ and $\bar{\aS}$ form a
representation under that bosonic algebra and obey
\begin{equation}
  \begin{aligned}
    \left[ \aM_{\mu\nu}, \aQ_{\alpha \check{\alpha}} \right]
    &= -\frac{1}{2} \left( \gamma_{\mu\nu} \aQ \right)
    _{\alpha \check{\alpha}}\,,\qquad&
    \left[ \aM_{\mu\nu}, \bar{\aS}_{\dot{\alpha} \check{\alpha}} \right]
    &= -\frac{1}{2} \left( \bar{\gamma}_{\mu\nu} \bar{\aS} \right)
    _{\dot{\alpha} \check{\alpha}}\,,\\
    \left[ \aK_\mu, \aQ_{\alpha \check{\alpha}} \right]
    &= \left( \gamma_\mu \bar{\aS} \right)
    _{\alpha \check{\alpha}}\,,&
    \left[ \aP_\mu, \bar{\aS}_{\dot{\alpha} \check{\alpha}} \right]
    &= \left( \bar{\gamma}_\mu \aQ \right)
    _{\dot{\alpha} \check{\alpha}}\,,\\
    \left[ \aD, \aQ_{\alpha \check{\alpha}} \right]
    &= \frac{1}{2} \aQ_{\alpha \check{\alpha}}\,, &
    \left[ \aD, \bar{\aS}_{\dot{\alpha} \check{\alpha}} \right]
    &= -\frac{1}{2} \bar{\aS}_{\dot{\alpha} \check{\alpha}}\,,\\
    \left[ \aR_{IJ}, \aQ_{\alpha \check{\alpha}} \right]
    &= \frac{1}{2} \left( \check{\gamma}_{IJ} \aQ \right)_{\alpha
      \check{\alpha}}\,,&
    \left[ \aR_{IJ}, \bar{\aS}_{\dot{\alpha} \check{\alpha}} \right]
    &= \frac{1}{2} \left( \check{\gamma}_{IJ} \bar{\aS} \right)
    _{\dot{\alpha} \check{\alpha}}\,.
  \end{aligned}
  \label{eqn:ospfermionrep}
\end{equation}
Finally, the anticommutator of $\aQ$ generates a translation
$\aP$, while the anticommutator of $\bar\aS$ generates a
special conformal transformation $\aK$
\begin{equation}
  \begin{gathered}
    \left\{ \aQ_{\alpha \check{\alpha}},
    \aQ_{\beta \check{\beta}} \right\}
    = 2 \left( \gamma_\mu c \right)_{\alpha \beta}
    \Omega_{\check{\alpha} \check{\beta}} \aP^\mu\,, \qquad\quad
    \left\{ \bar{\aS}_{\dot{\alpha} \check{\alpha}},
    \bar{\aS}_{\dot{\beta} \check{\beta}} \right\}
    = 2 \left( \bar{\gamma}_\mu c^T \right)_{\dot{\alpha} \dot{\beta}}
    \Omega_{\check{\alpha} \check{\beta}} \aK^\mu \,,\\
    \left\{ \aQ_{\alpha \check{\alpha}},
    \bar{\aS}_{\dot{\beta} \check{\beta}} \right\}
    = 2 \left[ \left(
      \aD
      + \frac{1}{2} \gamma_{\mu\nu} \aM^{\mu\nu}
      +\check{\gamma}_{IJ} \aR^{IJ}
      \right) c^T \Omega
  \right]_{\alpha \dot{\beta} \check{\alpha} \check{\beta}}\,.
  \end{gathered}
  \label{eqn:ospfermion}
\end{equation}
All the other commutators vanish.

Note that this algebra has a natural structure in terms of supermatrices.
This point of view, along with its relation to the 6d algebra presented above,
is elaborated in~\cite{claus:1997cq}. We also note that the $\sof(5)$ generators
can be expressed in terms of $\mathfrak{sp}(2)$ generators by the relation
\begin{align}
  \aU_{\check{\alpha} \check{\beta}} = \frac{1}{2} \left(
  \check{\gamma}_{I J} \Omega \right)_{\check{\alpha} \check{\beta}}
  \aR^{IJ}\,, \qquad
  \aR_{IJ} = -\frac{1}{4}
  \left( \Omega^\dagger \check{\gamma}_{IJ} \right)^{\check{\alpha} \check{\beta}}
  \aU_{\check{\alpha} \check{\beta}}\,.
\end{align}
The appropriate commutators are then
\begin{equation}
  \begin{gathered}
    \left[ \aU_{\check{\alpha} \check{\beta}},
    \aU_{\check{\gamma} \check{\delta}} \right]
    = 2 \Omega_{\check{\alpha} (\check{\gamma}} \aU_{\check{\delta}) \check{\beta}} 
    + 2 \Omega_{\check{\beta} (\check{\gamma}} \aU_{\check{\delta})
    \check{\alpha}}\,,\\
    \left[ \aU_{\check{\alpha} \check{\beta}},
    \aQ_{\alpha \check{\gamma}} \right]
    = 2 \aQ_{\alpha (\check{\alpha}} \Omega_{\check{\beta})
    \check{\gamma}}\,, \qquad
    \left[ \aU_{\check{\alpha} \check{\beta}},
    \bar{\aS}_{\dot{\alpha} \check{\gamma}} \right]
    = 2 \bar{\aS}_{\dot{\alpha} (\check{\alpha}}
    \Omega_{\check{\beta}) \check{\gamma}}\,.
  \end{gathered}
\end{equation}

\subsection{The subalgebra \texorpdfstring{$\osp{(4^*|2)} \oplus \osp{(4^*|2)}$}{osp(4|2) +
osp(4|2)}}
\label{sec:ospsubalg}

In the presence of the plane, the original symmetry $\osp{(8^*|4)}$ is reduced
to the subalgebra $\osp{(4^*|2)} \oplus \osp(4^*|2)$~\cite{dhoker:2008wvd}, a
real form of $D(2,1,\alpha) \oplus D(2,1,\alpha)$ with $\alpha = -1/2$. Each copy of the
$\osp{(4^*|2)}$ is a (rigid) 1d superconformal algebra, whose bosonic part is
\begin{equation}
  \begin{gathered}
  \left[ \aP_+, \aK_+ \right]
  = 2 \aD_+\,, \qquad
  \left[ \aD_+, \aP_+ \right] = \aP_+\,, \qquad
  \left[ \aD_+, \aK_+ \right] = -\aK_+\,,\\
  \left[ \aT_{(a)}^i, \aT_{(b)}^j \right]
  = -i \delta_{(ab)} \varepsilon^{ijk} \aT_{(b)}^k, \qquad
  (a) = 1,2.
  \end{gathered}
\end{equation}
In addition to the 1d conformal algebra, there are 2 additional $\su{(2)}$.
Together, they form the ``chiral'' part of the $\sof(2,2)_\parallel \oplus
\sof(4)_\perp \oplus \sof(4)_R$ preserved by the plane, with the ``antichiral''
part (denoted by a ``$-$'' subscript) given by the other $\osp{(4^*|2)}$. They are related
to the bulk generators by
\begin{align}
  \aP_\pm = \frac{1}{2} (\aP_0 \pm \aP_1)\,, \qquad
  \aD_\pm = \frac{1}{2} (\aD \pm \aM_{01})\,, \qquad
  \aK_\pm = \frac{1}{2}(-\aK_0 \pm \aK_1)\,,
\end{align}
where for definiteness we assume that the plane spans the directions $x^{0,1}$.
The decomposition of $\sof(4)_{\perp,R}$ is given by the 't Hooft symbols
\begin{align}
  \aT_{(1)}^{i_1} = \frac{i}{4} \eta^{i_1}_{ mn} \aM^{mn}\,, \qquad
  \aT_{(2)}^{i_2} = -\frac{i}{4} \eta^{i_2}_{ij} \aR^{ij}\,,
\end{align}
and similarly for $\bar{\aT}$ in terms of the antichiral 't Hooft symbols
$\bar{\eta}$.

In addition to these generators, the algebra includes supersymmetries
$\aQ_{\alpha_1 \alpha_2}$ and special supersymmetries $\aS_{\alpha_1 \alpha_2}$
charged under both $\su(2)$. These satisfy
\begin{equation}
  \begin{aligned}
    \left[ \aK_+, \aQ_{\alpha_1 \alpha_2} \right]
    &= -i \aS_{\alpha_1 \alpha_2}\,,&
    \left[ \aP_+, \aS_{\alpha_1 \alpha_2} \right]
    &= i \aQ_{\alpha_1 \alpha_2}\,,\\
    \left[ \aD_+, \aQ_{\alpha_1 \alpha_2} \right]
    &= \frac{1}{2} \aQ_{\alpha_1 \alpha_2}\,, &
    \left[ \aD_+, \aS_{\alpha_1 \alpha_2} \right]
    &= -\frac{1}{2} \aS_{\alpha_1 \alpha_2}\,,\\
    \left[ \aT_{(1)}^{i_1}, \aQ_{\alpha_1 \alpha_2} \right]
    &= \frac{1}{2} \tensor{(\sigma^{i_1})}{_{\alpha_1}^{\beta_1}}
    \aQ_{\beta_1 \alpha_2}\,,\qquad&
    \left[ \aT_{(1)}^{i_1}, \aS_{\alpha_1 \alpha_2} \right]
    &= \frac{1}{2} \tensor{(\sigma^{i_1})}{_{\alpha_1}^{\beta_1}}
    \aS_{\beta_1 \alpha_2}\,,\\
    \left[ \aT_{(2)}^{i_2}, \aQ_{\alpha_1 \alpha_2} \right]
    &= \frac{1}{2} \tensor{(\sigma^{i_2})}{_{\alpha_2}^{\beta_2}}
    \aQ_{\alpha_1 \beta_2}\,,&
    \left[ \aT_{(2)}^{i_2}, \aS_{\alpha_1 \alpha_2} \right]
    &= \frac{1}{2} \tensor{(\sigma^{i_2})}{_{\alpha_2}^{\beta_2}}
    \aS_{\alpha_1 \beta_2}\,,
  \end{aligned}
\end{equation}
where $\sigma^{i}$ are the Pauli matrices. They anticommute to
\begin{equation}
  \label{eqn:osp4susy}
  \begin{gathered}
    \left\{ \aQ_{\alpha_1 \alpha_2},
    \aQ_{\beta_1 \beta_2} \right\}
    = 2 i \epsilon_{\alpha_1 \beta_1} \epsilon_{\alpha_2 \beta_2}
    \aP_+\,, \qquad\quad
    \left\{ \aS_{\alpha_1 \alpha_2},
    \aS_{\beta_1 \beta_2} \right\}
    = 2 i \epsilon_{\alpha_1 \beta_1} \epsilon_{\alpha_2 \beta_2}
    \aK_+\,,\\
    \left\{ \aQ_{\alpha_1 \alpha_2},
    \aS_{\beta_1 \beta_2} \right\}
    = 2 \left[
       \epsilon_{\alpha_1 \beta_1} \epsilon_{\alpha_2 \beta_2} \aD_+
      + (\sigma^{i_1} \epsilon)_{\alpha_1 \beta_1} \epsilon_{\alpha_2 \beta_2}
      \aT_{(1)}^{i_1}
      - 2 \epsilon_{\alpha_1 \beta_1} (\sigma^{i_2} \epsilon)_{\alpha_2 \beta_2}
      \aT_{(2)}^{i_2}
  \right]\,.
  \end{gathered}
\end{equation}
The ratio $\alpha = -1/2$ between the coefficients of $\aT_{(1)}$ and
$\aT_{(2)}$ is a specific case of the exceptional Lie algebra $D(2,1;\alpha)$
(see~\cite{Sevrin:1988ew} for the algebra with general $\alpha$ and its
Kac-Moody extension).

The precise embedding of these supercharges inside $\aQ_{\alpha \check{\alpha}}$
is obtained by restricting to the preserved supercharges $\Pi_+ \aQ = \aQ$,
where the projector is~\cite{Drukker:2020dcz}
\begin{align}
  \tensor{(\Pi_\pm)}{_{\alpha \check{\alpha}}^{\beta \check{\beta}}}
  = \frac{1}{2} \tensor{\left[ 1 \pm \gamma_{01} \check{\gamma}_5 \right]}
    {_{\alpha \check{\alpha}}^{\beta \check{\beta}}}\,, \qquad
    \tensor{(\Pi_\pm)}{_{\dot{\alpha} \check{\alpha}}^{\dot{\beta} \check{\beta}}}
    = \frac{1}{2} \tensor{\left[ 1 \mp \bar{\gamma}_{01} \check{\gamma}_5 \right]}
    {_{\dot{\alpha} \check{\alpha}}^{\dot{\beta} \check{\beta}}}\,,
  \label{eqn:ospprojector}
\end{align}
which has a different expression acting respectively on chiral and antichiral
representations. This projector decomposes as
\begin{align}
  \frac{1}{2} \left[ 1 + \gamma_{01} \check{\gamma}_5 \right] =
  \frac{1}{2} \left[ 1 + \gamma_{01} \right]
  \frac{1}{2} \left[ 1 + \check{\gamma}_5 \right]
  + \frac{1}{2} \left[ 1 - \gamma_{01} \right]
  \frac{1}{2} \left[ 1 - \check{\gamma}_5 \right],
\end{align}
which gives, respectively for the two terms, two anticommuting supercharges
$\bar{\aQ}_{\dot{\alpha}_1 \dot{\alpha}_2}$ and $\aQ_{\alpha_1 \alpha_2}$. Their
chirality is derived from the projector: $(1 + \gamma_{01})$ projects onto the
positive chirality component, which is correlated with the positive chirality
under $\sof(4)_\perp$ since $\gamma_{01} = \gamma_{2345}$.

\subsubsection{Subalgebra as an embedding inside
\texorpdfstring{$\osp{(8^*|4)}$}{osp(8|4)}}
\label{sec:repderivation}

Lastly, in Section~\ref{sec:disp} and~\ref{sec:BulkDefectCorrelators} it is
convenient to discuss the subalgebra directly within the larger $\osp(8^*|4)$.
Here we decompose some of the commutators of $\osp(8^*|4)$ into preserved and broken
generators directly with the projector. We make use of the following identities
\begin{gather}
  \nonumber \Pi_\pm^\dagger = \Pi_\pm, \quad
  (\Pi_\pm \mathcal{C})^T = -\Pi_\pm \mathcal{C}^T\,,\\
  \label{eqn:ospprojp}
  \left[ \Pi_\pm, \Gamma_a \right] =
  \left[ \Pi_\pm, \check{\gamma}_5 \right] = 0\,,\\
  \nonumber \Pi_\pm \Gamma_{m} = \Gamma_{m} \Pi_\mp, \quad
  \Pi_\pm \Gamma_{i} = \Gamma_{i} \Pi_\mp\,.
\end{gather}
Note that here we don't differentiate between the action of $\aQ$ and
$\bar{\aQ}$ for simplicity.

Using these properties, one can easily derive the induced subalgebra and its
representation by acting with $\Pi_\pm$.  The only nontrivial part of the
preserved algebra is for the supercharges, which now obey
\begin{equation}
  \begin{gathered}
  \left\{ \aQ^+_{ \alpha \check{\alpha}},
  \aQ^+_{ \beta \check{\beta}} \right\}
  = 2 \left( \gamma_a \Pi_+ c \Omega \right)_{\alpha
  \beta \check{\alpha} \check{\beta}}
  \aP^a\,,\qquad\quad
  \left\{ \bar{\aS}^+_{\dot{\alpha} \check{\alpha}},
  \bar{\aS}^+_{\dot{\beta} \check{\beta}} \right\}
  = 2 \left( \bar{\gamma}_a \Pi_+ c^T \Omega 
  \right)_{\dot{\alpha} \dot{\beta} \check{\alpha} \check{\beta}}
  \aK^a\,,\\
  \left\{\aQ^+_{\alpha \check{\alpha}},
  \bar{\aS}^+_{\dot{\beta} \check{\beta}} \right\}
  = 2 \left[ \left( \check{\gamma}_{ij} \aR^{ij}
    + \aD
    + \frac{1}{2} \gamma_{mn} \aM^{mn}
    + \frac{1}{2} \gamma_{ab} \aM^{ab}
  \right) \Pi_+ c^T \Omega \right]_{\alpha \dot{\beta} \check{\alpha}
  \check{\beta}}\,.
  \end{gathered}
\end{equation}
The broken generators satisfy

\begin{minipage}[b]{0.25\textwidth}
  \begin{tikzpicture}[scale=2.4,baseline]
    \node (np) at (0,1) {$\aP_{m}$};
    \node (nq) at (0.5,0.5) {$\aQ^-$};
    \node (nr) at (1,0) {$\aR_{i 5}$};
    \node (nm) at (0,0) {$\aM_{a m}$};
    \node (ns) at (0.5,-0.5) {$\bar{\aS}^-$};
    \node (nk) at (0,-1) {$\aK_{m}$};
    \draw[thick,->] (nk) edge node[right,anchor=north west] {$\aQ_+$} (ns) (ns) edge (nr) (nr) edge (nq) (nq) edge (np)
    (ns) edge (nm) (nm) edge (nq);
    \draw[dashed,->] (nk) edge node[left,anchor=east] {$\aP_a$} (nm)
    (nm) edge (np) (ns) edge (nq);
  \end{tikzpicture}
\end{minipage}
\begin{minipage}[b]{0.74\textwidth}
\begin{equation}
  \begin{aligned}
    \left[\aQ^+_{\alpha \check{\alpha}},
    \aP_{m} \right] &= 0,\\
    \left\{\aQ^+_{\alpha \check{\alpha}},
     \aQ^-_{\beta \check{\beta}} \right\}
    &= 2 \left( \gamma_{m} \Pi_- c \Omega \right)
    _{\alpha \beta \check{\alpha} \check{\beta}} \aP^{m},\\
    \left[ \aQ^+_{\alpha \check{\alpha}},
    \aR_{i 5} \right]
    &= -\frac{1}{2} \left( \check{\gamma}_{i 5} \aQ^-
    \right)_{\alpha \check{\alpha}},\\
    \left[ \aQ^+_{\alpha \check{\alpha}},
    \aM_{a m} \right]
    &= \frac{1}{2} \left( \gamma_{a m} \aQ^-
    \right)_{\alpha \check{\alpha}},\\
    \left\{ \aQ^+_{\alpha \check{\alpha}},
    \bar{\aS}^-_{\dot{\beta} \check{\beta}} \right\}
    &= 4\left[ \left( 
      \check{\gamma}_{i 5} \aR^{i 5}
      + \frac{1}{2} \gamma_{a m} \aM^{a m} \right)
    \Pi_- c^T \Omega \right]_{\alpha \dot{\beta} \check{\alpha} \check{\beta}},\\
    \left[ \aQ^+_{\alpha \check{\alpha}},
    \aK_{m} \right]
    &= -\left( \gamma_{m} \bar{\aS}^- \right)
    _{\alpha \check{\alpha}}.
  \end{aligned}
  \label{eqn:brokenalg}
\end{equation}
\end{minipage}
\\

\noindent These transformations are related to~\eqref{eqn:dispsusy}
using~\eqref{eqn:broken} to write the displacement operator as contact terms in
the presence of the defect:
\begin{align}
  \aR^{i5} V =
  \int_{\mathbb{R}^2} \diff^2 \sigma V[\bO^{i}(\sigma)]\,.
\end{align}
We can recover the full representation by acting with $\aQ^+$, e.g.,
\begin{align}
  \int_{\mathbb{R}^2} V[ \aQ^+ \bO^{i}(\sigma) ]
  \diff^2 \sigma =
  \left[ \aQ^+, \aR^{i5} \right] V 
  = -\frac{1}{2} \check{\gamma}_{i5} \aQ^- V
  = -\frac{1}{2} \int_{\mathbb{R}^2} \diff^2 \sigma
  V[\check{\gamma}_{i5} \mathbb{Q}^-(\sigma) ]\,.
\end{align}
The action of $\aQ^+$ on $\mathbb{Q}$ can similarly be read
from~\eqref{eqn:brokenalg}, but it misses the descendant. These are fixed
instead by requiring closure under the Jacobi identity as in~\eqref{eqn:jacobi}
(see also for instance the discussion in Section 2 of~\cite{Bianchi:2020hsz}).

\bibliographystyle{utphys2}
\bibliography{ref}

\end{document}